\documentclass[12pt]{article}
\pdfoutput=1

\usepackage{graphics, color}
\usepackage{graphicx}
\usepackage{amssymb}

\def\gsim{\, \rlap{$>$}{\lower 1.1ex\hbox{$\sim$}}\,}
\def\lsim{\, \rlap{$<$}{\lower 1.1ex\hbox{$\sim$}}\,}

\newcommand{\be}{\begin{equation}}
\newcommand{\ee}{\end{equation}}
\newcommand{\bea}{\begin{eqnarray}}
\newcommand{\eea}{\end{eqnarray}}

\begin{document}


\begin{titlepage}

\setcounter{page}{1} \baselineskip=15.5pt \thispagestyle{empty}

\vbox{\baselineskip14pt
}
{~~~~~~~~~~~~~~~~~~~~~~~~~~~~~~~~~~~~
~~~~~~~~~~~~~~~~~~~~~~~~~~~~~~~~~~
~~~~~~~~~~~ \footnotesize{SU/ITP-14/02, SLAC-PUB-15905}} \date{}

\bigskip\

\vspace{.5cm}
\begin{center}
{\fontsize{19}{36}\selectfont  \sc
 Backdraft:  String Creation in an old Schwarzschild Black Hole\\ 
\vspace{5mm}
}
\end{center}


\vspace{0.6cm}

\begin{center}
{\fontsize{13}{30}\selectfont  Eva Silverstein}
\end{center}


\begin{center}
\vskip 8pt
\textsl{
Stanford Institute for Theoretical Physics, Stanford University, Stanford, CA 94306, USA}

\vskip 7pt
\textsl{ SLAC National Accelerator Laboratory, 2575 Sand Hill, Menlo Park, CA 94025}



\end{center}

\vspace{0.2cm}
\hrule \vspace{0.1cm}
{ \noindent \textbf{Abstract} \\[0.2cm]
\noindent 

We analyze string production in the background of a Schwarzschild black hole, after developing first quantized methods which capture string-theoretic nonadiabatic effects which can exceed naive extrapolations of effective field theory.    Late-time infalling observers are strongly boosted in the near horizon region relative to early observers and formation matter.  In the presence of large boosts in flat spacetime, known string and D-brane scattering processes exhibit enhanced string production, even for large impact parameter.  This suggests the possibility that the nonadiabatic dynamics required to realize the firewall proposal of AMPS occurs for old black holes, with the late-time observer catalyzing the effect.  
After setting up this dynamical thought experiment,
we focus on a specific case: the production of open strings stretched D-particles, at least one of which falls in late (playing the role of a late time observer).  For relatively boosted D-branes, we precisely  recover earlier results of Bachas, McAllister and Mitra which we generalize to brane trajectories in the black hole geometry.   For two classes of late-time probes, we find a regime of significant non-adiabaticity by horizon crossing, assessing its dependence on the boost in each case.  Closed string probes, as well as additional effects in D-brane scattering, may produce other significant non-adiabatic effects depending on the boost, something we leave for further work.  

}

 \vspace{0.3cm}
 \hrule

\vspace{0.6cm}
\end{titlepage}

\tableofcontents

\newpage

\baselineskip = 16pt

\section{Introduction, setup, and overview}

In this paper, we study non-adiabatic string-theoretic dynamics in the presence of a late-time probe of a Schwarzchild black hole, developing general methods for computing string creation generalizing \cite{Bachas}\cite{RelDbranes}\cite{CSGW}\cite{Gubser}\cite{AlbionEmil}\cite{Strominger}.  We explore implications for the AMPS paradox \cite{AMPS}\ (and its precursors such as \cite{preAMPS}), investigating string production in the presence of a relative boost as a dynamical mechanism for generating a firewall, an effect which can be catalyzed by the late-time observer itself.  In addition, string production may provide dynamical limits on certain thought experiments. 


Although the paradox \cite{AMPS}\ could be derived using the low energy effective theory \cite{hawking}\cite{mathurtheorem}, the resulting firewall proposal is sensitive to the UV completion, so it is reasonable to seek a resolution at that level.\footnote{One approach in this general spirit is the fuzzball program \cite{mathurtheorem}\cite{mathurprogram}, which involves very different aspects of string theory from those we consider here, as do the approaches \cite{otherstringy}\cite{stringBH}.}  
String theory provides a well studied candidate for the ultraviolet completion of gravity, and underlies the explicit AdS/CFT systems for which unitarity follows from the duality \cite{AdSCFT}.  String-theoretic  effects go beyond effective field theory, in some known cases leading to significantly stronger non-adiabatic effects than would be obtained by a na\"ive extrapolation of effective field theory \cite{Bachas}\cite{RelDbranes}\cite{CSGW}\cite{Veneziano}.  

Many inventive ideas for resolving the AMPS paradox have emerged in recent months, including discussion of radical modifications of quantum theory.  However, although \cite{AMPS}\ carefully showed that effective field theory is not consistent with other postulates in black hole physics,  there has not been a similar analysis at the level of string theory.
Assessing this requires analyzing non-adiabatic effects more thoroughly in string theory (or more general UV completions of gravity), to understand clearly where the effective field theory vacuum should break down in that framework.   A related motivation for analyzing the black hole horizon within a UV completion of gravity is the appearance of a ``trans-Planckian problem" in some putative processes involving mode propagation over long times in an old black hole geometry.  In particular, late-time modes just inside the horizon which propagate up from the formation matter undergo a large blueshift.      
Relatedly, as we will review shortly, in the locally flat near-horizon region of the black hole, late-time probes are highly boosted relative to 
early infallers (and the formation matter).  

This relative boost especially motivates a careful analysis of string-theoretic non-adiabatic effects in black hole physics.    
In string theory, large boosts can lead to strong non-adiabatic effects, as studied in other contexts.  Examples include open string pair production on D-branes \cite{Bachas}\cite{RelDbranes}, and multiple closed-string production in high energy string scattering \cite{Veneziano}\ as well as tidal-force-induced excitations on scattered strings.  These effects grow with boost parameter $\eta = {\rm arctanh}({\rm velocity})$, with the impact parameter $b$ at which the effect remains significant growing linearly \cite{Bachas}\cite{RelDbranes}\  or exponentially \cite{Veneziano}\cite{RelDbranes}\ as a function of $\eta$ depending on the process involved.   
We would like to understand if analogous effects occur in the black hole problem as a result of the large boost in the trajectory of a late-time observer.  

More generally, we would simply like to develop reliable techniques for computing non-adiabatic effects in string theory. 
This question is partly motivated by the need for a clear dynamical assessment of the firewall proposal \cite{AMPS}, but may have other applications.   Just at the level of pair production in time dependent systems, various previous examples such as \cite{Bachas}\cite{RelDbranes}\cite{CSGW}\ exhibit stronger non-adiabaticity than would arise from a naive extrapolation of particle production, obtained by simply replacing the particle spectrum by the flat spacetime single-string oscillator spectrum.  This happens for clear physical reasons in each case, and goes beyond effective field theory.
Estimating non-adiabaticity by considering strings stretched along spacelike slices in a geometry is somewhat subtle, as the results cannot depend on the time slicing involved in such a picture.  We will develop a first-quantized path integral method for computing string pair production, and test it against a variety of known examples before applying it to black hole physics.\footnote{Earlier works addressing string-theoretic effects in black hole physics include \cite{stringBH}, and it would be interesting to understand if there is any connection, or application of that previous work to our two-probe thought experiment.}  

\begin{figure}[htbp]
\begin{center}
\includegraphics[width=10cm]{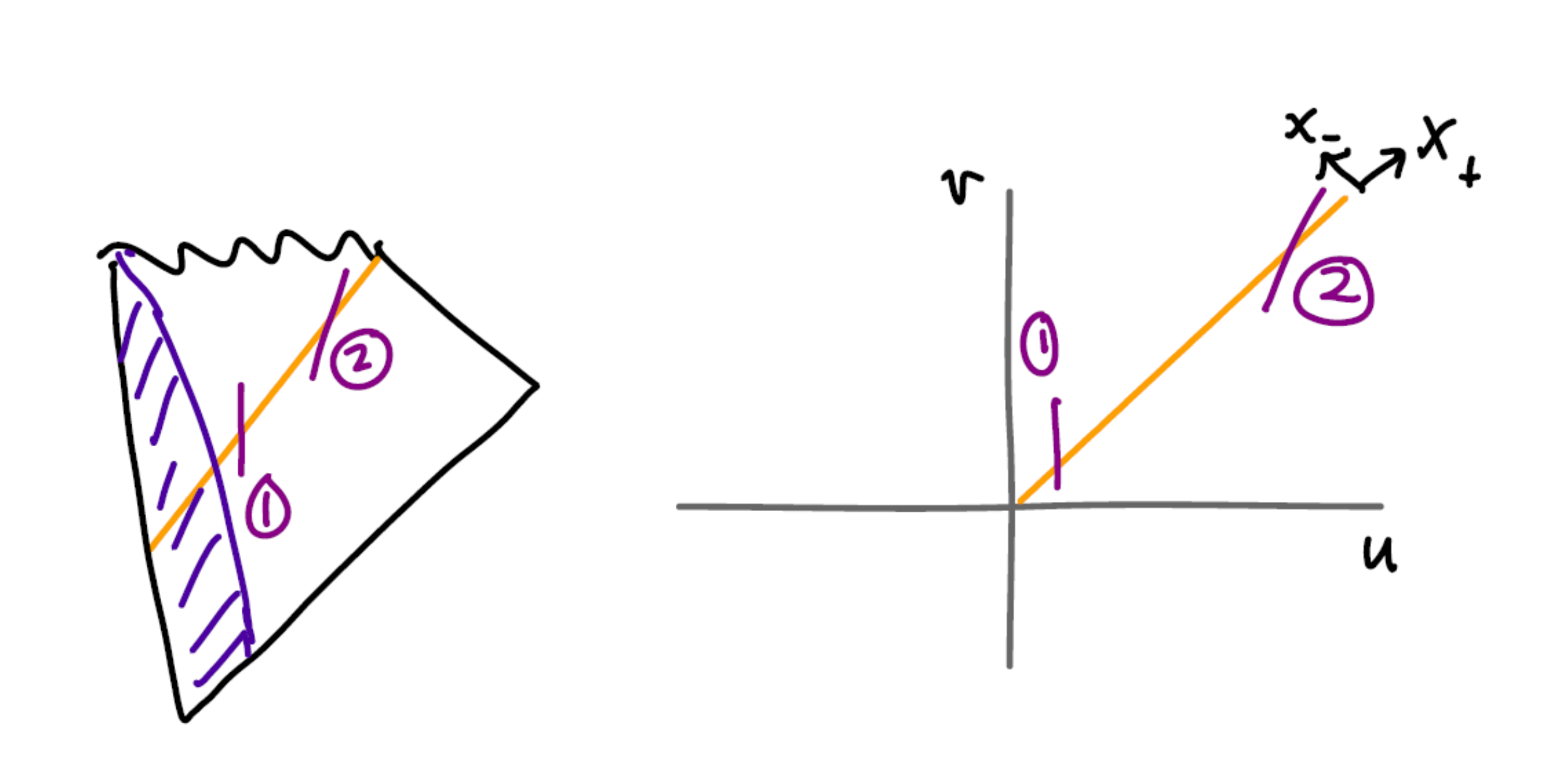}
\end{center}
\caption{The relatively boosted trajectories 1 and 2 described in the text.  Near the horizon, the spacetime reduces to a patch of Minkowski space. This raises the question of whether the strong boost-induced non-adiabatic effects described in \cite{Bachas}\cite{RelDbranes}\cite{Veneziano}\ might arise for probes of black holes, providing a dynamical mechanism for the breakdown of effective field theory.}
\label{boosttrajectories}
\end{figure}

Let us start by briefly formulating a class of dynamical thought experiments which involve near-horizon boost, before moving on to the specific cases analyzed in this paper.
Consider a string or D-brane sent into a Schwarzschild black hole
\be\label{Schwarz}
ds^2 = -(1-\frac{r_0}{r})dt^2 + \frac{dr^2}{(1-\frac{r_0}{r})}+r^2 d\Omega^2.
\ee 
As we will review in detail shortly, if we compare an early probe (or formation matter) to a late probe -- by shifting two trajectories relative to each other by a Schwarzschild time translation $\Delta t$ -- the two are relatively boosted by $\eta=\Delta t/2r_0$ as they cross the horizon (see figure \ref{boosttrajectories}).\footnote{Another way to get a strongly boosted trajectory in the near horizon region, starting from a fixed time $t_0$, is to drop the particle in from further away, changing the ratio $E/m$ of the particle energy to its mass.  We will find it to be a useful technical simplification to use the translation $\Delta t$ along a direction of symmetry to introduce our late probe, introducing it with the same value of $E/m$ as the early probe.}  We would like to understand if large string-theoretic non-adiabatic effects arise, and how they depend on the large boost between the early and late probes.           

It is difficult to obtain an immediate estimate for the production of strings in this system.  Intuitively, strings extended between the probes are stretched apart by them, suggesting a growing mass which can lead to non-adiabatic production.  In particular, in the cosmological geometry of the black hole interior, the proper length of a string stretched from $t$ to $t+\Delta t$ (along slices of constant $r$) grows rapidly with time.  But the proper distance between probes depends on the spatial slices, something that cannot affect physical results.  

In order to obtain reliable estimates for the level of production, we will develop a semiclassical worldsheet path integral method for computing string pair production which is particularly practical to use in symmetric configurations.  We will recover earlier results along the way such as \cite{CSGW}\cite{Bachas}\cite{RelDbranes}, and this method may be useful more generally.  After warming up with a number of examples, we will apply our method to open strings on D-branes in the black hole, treated as a background for the strings.  We will find two families of probes for which there is significant non-adiabatic production by the time of horizon crossing.  
The level of non-adiabaticity depends on the ratio $C=E/m$ of the D-particle energy to its mass.\footnote{As we will discuss below, this is consistent with parametrically sub-Planckian probe energy.}  For $C<1$ in (\ref{ImSBH})  we find boost-enhanced non-adiabaticity which is related to an early brane interaction in the corresponding trajectories, although the region of non-adiabaticity is not limited to the location of the near-collision. 
This may provide an interesting dynamical limitation on the corresponding thought experiment. 
Moreover, for very small $C$, $C\ll\sqrt{\alpha'}/r_0$, we find strong and boost-enhanced non-adiabaticity at the horizon, according to an estimate based on the time variation of the frequency in an infalling frame in equation (\ref{boostenhanced}).  

Once pairs of open strings are produced, in the regime of parameters for which this happens, they decay into closed strings and radiation, leading to a rich excited state above the effective field theory vacuum.\footnote{This pair production of open strings on D-branes is just the simplest string theoretic process to consider, which dominates in some controlled regimes.   But we reiterate that there are additional effects and other cases to check, which may give qualitatively different results.  In particular,  in studies of string and brane scattering processes in Minkowski spacetime, various effects including Bremsstrahlung give stronger, exponential, behavior as a function of the boost parameter $\eta$ \cite{Veneziano}\cite{RelDbranes}\cite{stringBH}.   Specifically, at an impact parameter growing like $e^\eta$, the works \cite{Veneziano}\ find strong inelastic effects, and the second reference of \cite{RelDbranes}\ finds significant Bremsstrahlung.   The distinction between linear and exponential behavior with boost parameter $\Delta t/r_0$ is reminiscent of the distinction between the Page time \cite{Pagetime}\ and the scrambling time \cite{scrambling}.  We thank Thomas Bachlechner,  Matthew Dodelson, Steve Giddings, Liam McAllister, Gabriele Veneziano, and Danjie Wenren for interesting discussions along these lines, and leave it for future work.}     

We also estimate the level of non-adiabaticity near the horizon for more general $C$ in Painlev\'e coordinates , again via its relation to the time variation of the effective frequency.  This corresponds to view of auxiliary observers dropped in from rest at infinity.   The resulting estimate for the total number of produced pairs of open strings (for sufficiently large $C$) is given in equation (\ref{Ntotgen}) below, and for more general $C>1$ it follows from the formula (\ref{nonadGPresult}).  There is nontrivial dependence on the boost parameter $\eta=\Delta t/2 r_0$.  In this case, although in some regimes of parameters the effect grows with boost, in the regime we find (\ref{Ntotgen}) of strongest non-adiabaticity this is not the case and the relative boost in fact suppresses the effect.  However, in this case there is a strong boost between the auxiliary Painlev\'e observers and the infalling probe.

Technically, the basic strategy is to derive the leading saddle point contribution to the production amplitude,
using the symmetries of the problem to simplify the analysis.
In field theory the required amplitude is related to a component of the two-point correlation function.  We rederive this using a first quantized method which generalizes to the worldsheet path integral in string theory.  
Again, there are simple regimes controlled by WKB in which the latter does not reduce to a naive extrapolation of particle production, a phenomenon noted in \cite{Bachas}\cite{RelDbranes}\cite{CSGW}.  In most of our calculations, we explicitly evaluate the saddle point corresponding to particle or string production.  In the  trajectories in the black hole, we also make a related but cruder real-time estimate of the near horizon non-adiabaticity, both in an infalling frame and in Painleve coordinates.  This is the only estimate we make in the case of $C=E/m >1$,  since in that case the effect continues to grow further inside the horizon of the black hole and we do not wish to count production coming from further in near the singularity.      

There are inherent uncertainties in finite-time production calculations, a caveat that is especially significant in the case of string theory whose precise observables in flat spacetime are limited to S-matrix elements.  We nonetheless expect that our calculation provides a good approximation to the physics and could in principle be extracted from a complete S-matrix treatment, but we will not accomplish that in the present work.    

For comparison and as a check, we will obtain approximately adiabatic results for some other cases.  These include 
open string pair production between unboosted D-branes and between D-branes in the special case of BTZ \cite{BTZ}\ and topological black holes \cite{emparanhyp}\ at the mass level for which they are related to pure $AdS$ spacetime via an orbifold of the system on its Coulomb branch \cite{InsightfulD}.  We will also similarly analyze the simplest analogous probe in the de Sitter static patch.

\subsection{Previous results on relatively boosted branes in Minkowski spacetime}

Let us recall in more detail the result of \cite{Bachas}\cite{RelDbranes}\ on the production of pairs of open strings ending on relatively relativistically boosted D$0$-branes in Minkowski spacetime,\footnote{We will reproduce the results of \cite{Bachas}\cite{RelDbranes}\ below using our methods; here we simply quote their results.} with velocity
\be\label{vel}
V_{relative}=\tanh\eta, ~~~~ \eta\gg 1.
\ee
and impact parameter $b$.
The number of produced open strings behaves as
\be\label{Ntot}
\sum_n \rho(n) e^{-\frac{\pi b^2}{\alpha'\eta}-\frac{2\pi^2 n}{\eta}}
\ee
where $\alpha'$ is the inverse string tension and $\rho(n)$ is the number of states at oscillator level $n$.
The effect is strongly enhanced at large boost $\eta\gg 1$.   

In particular, the na\"ive extrapolation of effective field theory would give a result depending on the relative velocity $V_{rel}=\tanh\eta$ in place of $\eta$ in (\ref{Ntot}).   
That is, the mass of a string stretched between branes along a Minkowski spatial slice is $\sqrt{V_{rel}t_{m}^2+b_\perp^2}/\alpha'$, where $t_m$ is the Minkowski time.  The Bogoliubov coefficient $\beta$ describing the mixing between positive and negative frequency modes in the presence of this time dependent mass satisfies $|\beta|^2=e^{-\frac{\pi b^2}{\alpha'V_{rel}}}$.  This would be a gross underestimate of the actual effect, which depends directly on $\eta$ as in (\ref{Ntot}).  In that sense, the example \cite{Bachas}\cite{RelDbranes}\ is enhanced relative to a standard effective field theory estimate.  Nonetheless the effect is standard in string theory.   

We will recover this result and that of \cite{CSGW}\ precisely below with our methods.
As described in the work \cite{Bachas}\cite{RelDbranes}, this effect can be understood qualitatively as follows using some features of string theory.  The open string spectrum contributing to the annulus diagram 
exhibits an effective reduction in the string tension, related by T-duality to the 
analogous effect near a critical electric field, in which the effective tension is suppressed as the field pushes apart the string endpoints.   It is useful to fold in the density of string states, $\rho(n)\sim e^{\sqrt{8\pi^2 n}}$.  From (\ref{Ntot}), for the peak value $\sqrt{n_{peak}}=\eta/\sqrt{2\pi^2}$ we obtain a factor of $e^\eta$ from the sum over oscillator levels.  As a result, one finds significant open string production even at an impact parameter $b$ as large as $b_{open}\sim\eta\sqrt{\alpha'}$.   

There are several interesting regimes of brane and string scattering \cite{Veneziano}\cite{RelDbranes}, some exhibiting distinct effects such as strong gravitational dynamics or Bremsstrahlung.  The open string production effect discussed here dominates over strong gravitational effects in a controlled regime where $b_{open}$ exceeds the Schwarzschild  radius corresponding to the energy of the scattering D-branes.  Bremsstrahlung, while very interesting, can be avoided by setting up a state of relative relativistic motion at the time of onset of open string non-adiabaticity.\footnote{We thank S. Giddings, T. Bachlechner, and L. McAllister for discussions.}  In the present work, we will focus on the calculation of open string pair production, as the diagrammatically simplest effect, leaving other regimes and other contributions for further work.  

\subsection{Black hole trajectories and boosts}

The relative boost of interest in black hole physics between early matter and late probes arises from Schwarzschild time translation as follows.  It can be seen most simply by writing the Schwarzschild black hole (\ref{Schwarz})      
in  Kruskal-Szekeres coordinates
\bea\label{KScoords}
v &=& r_0\sqrt{\frac{r}{r_0}-1}e^{r/2r_0} \sinh(\frac{t}{2r_0}) , ~~~ u =  r_0\sqrt{\frac{r}{r_0}-1}e^{r/2r_0} \cosh(\frac{t}{2r_0}) ~~~~~ {(r>r_0)}   \\
v &=&  r_0\sqrt{1-\frac{r}{r_0}}e^{r/2r_0} \cosh(\frac{t}{2r_0}), ~~~ u = r_0 \sqrt{1-\frac{r}{r_0}}e^{r/2r_0} \sinh(\frac{t}{2r_0})  ~~~~~ {(r<r_0)}    
\eea
with metric
\be\label{Kruskal}
ds^2=\frac{4 r_0}{r} e^{-r/r_0}\left(-dv^2+du^2\right)+r^2d\Omega^2.
\ee
Near the horizon $r\approx r_0$, this describes a patch of Minkowski space.  For $r<r_0$, it is a Milne-like region in the $r, t$ directions, and for $r>r_0$ it is a Rindler-like region.  

We can see from (\ref{KScoords}) that shifting a trajectory by $\Delta t$ produces a second trajectory which near the horizon is boosted relative to the first one by a factor
\be\label{boost}
e^\eta \equiv e^{\Delta t/2r_0}
\ee  
This is perhaps simplest to state in light cone coordinates
\be\label{XpmUV}
x_+=v+u, ~~~~ x_-=v-u.
\ee
which transform as
\be\label{boostxpm}
x_\pm \to e^{\pm\eta} x_\pm.
\ee
under  a boost  (\ref{boost}).  

\begin{figure}[htbp]
\begin{center}
\includegraphics[width=8cm]{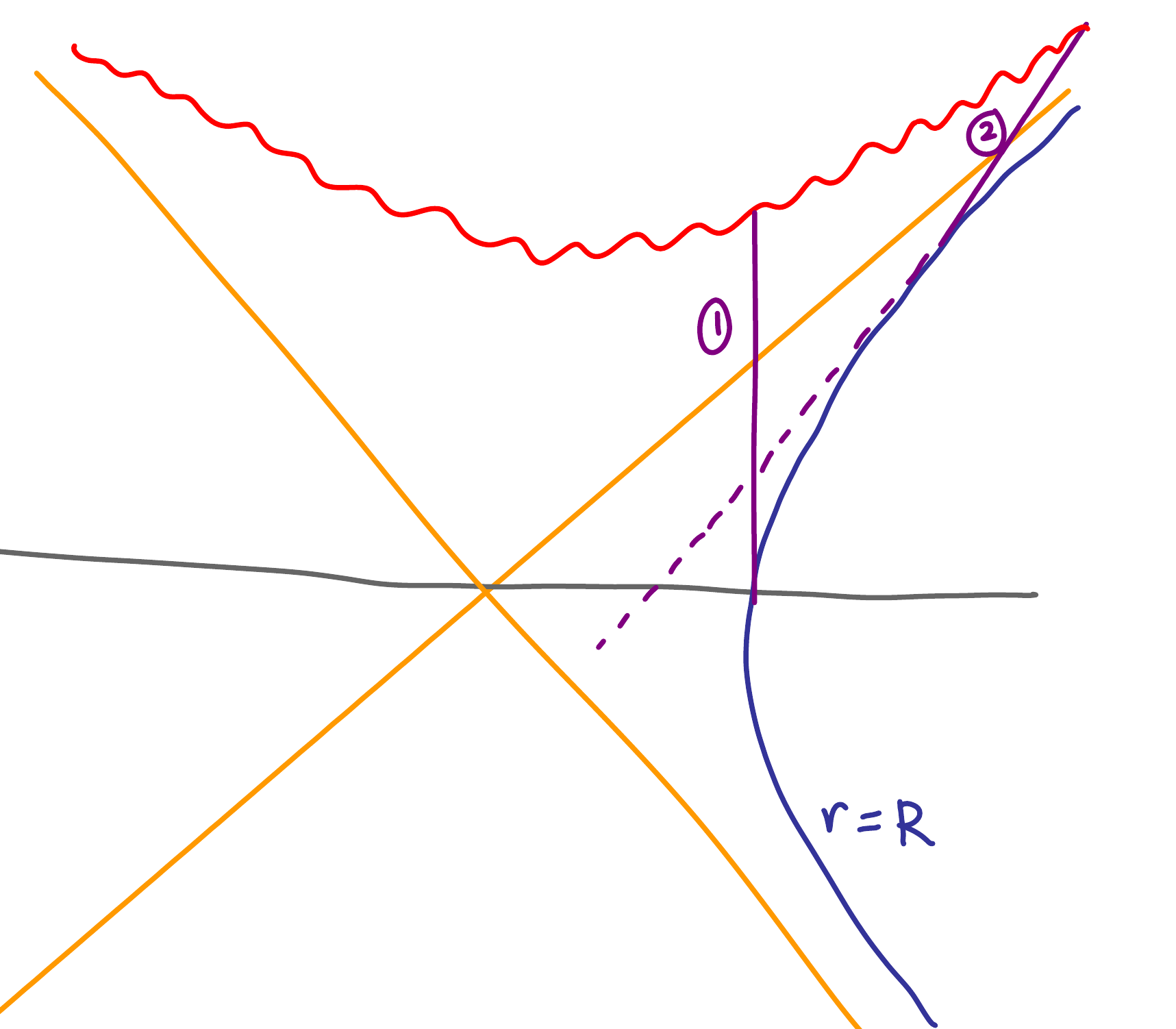}
\end{center}
\caption{A Kruskal diagram of the setup for $E<m$.  Starting from a constant radial position $r=R$ outside the black hole, two D-branes are dropped in at very different times $t=0$ and $t=\Delta t$.  As we increase $\Delta t$, the intersection point between the solid and dashed purple lines approaches the horizon as in (\ref{meeting}).}
\label{SetupKruskal}
\end{figure}

The path of a given particle near the horizon also depends on its energy and angular momentum.
For simplicity, let us consider a radially infalling particle.   The trajectory depends on the ratio $E/m$, with $E$ the conserved energy of the trajectory 
\be\label{rEM}
E=m\frac{\sqrt{1-r_0/r}}{\sqrt{1-(\frac{dr}{dt})^2/(1-r_0/r)^2}}.
\ee
For $E<m$, the particle falls from rest at a radial position 
\be\label{REm}
r=R=\frac{r_0}{1-E^2/m^2}, 
\ee
In the eternal black hole geometry, this is a turning point it reaches after moving outward from the past horizon.   Two such trajectories displaced by $\Delta t$ cross paths as the first falls in while the second is still moving radially outward (see figure \ref{SetupKruskal}).  For $E=m$, the particle falls in from rest at $r=R=\infty$.  
For $E>m$, the particle is sent into the black hole with nonzero velocity. 

We will be interested in these various families of trajectories in our string production calculations, with two D-particles following a trajectory of a given $E/m$, displaced by $\Delta t$ in Schwarzschild time.  We will warm up with the $E<m$ case, where we will find an extra saddle point contribution to the worldsheet
path integral which describes production triggered by the early collision of the two trajectories (up to a transverse impact parameter), see figure \ref{SetupKruskal}.  The spacetime region in which the non-adiabaticity originates, as well as its amplitude, will depend on $E/m$.  The non-adiabatic production in this case is enhanced at large boost $\eta=\Delta t/2r_0$.  For sufficiently small $E/m$, we will find non-adiabaticity at the horizon which is boost-enhanced.    

In the cases with $E\ge m$, there is no early collision of the two trajectories, and in that case we will also ultimately focus on isolating effects arising near horizon crossing. 
In this case, we will find significant open string pair production near the horizon for sufficiently large $E/m$, from the point of view of auxiliary observers dropped in from infinity.

In more detail, the $E<m$ trajectories can be expressed as \cite{MTW}\cite{trajpaper}\
\bea\label{radialfall}
r(\alpha) &=& \frac{R}{2}(1+\cos(\alpha))\\ \nonumber
t(\alpha) &=&( (\frac{R}{2}+r_0)\alpha+\frac{R}{2}\sin(\alpha))\sqrt{\frac{R}{r_0}-1}+r_0 \log\left(\left|\frac{\sqrt{\frac{R}{r_0}-1}+\tan(\alpha/2)}{\sqrt{\frac{R}{r_0}-1}-\tan(\alpha/2)}\right|\right)\\ \nonumber
\eea
with the parameter $\alpha$ increasing from $0$ to $\pi$ as $r$ decreases from $R$ to $0$.  
Given the first solution, where $r(0)=R, t(0)=0$, the $t$ translation symmetry determines the second solution to simply be given by shifting $t\to t+\Delta t$.   


The trajectories with $E\ge m$ satisfy
\bea\label{tofrintro}
t(r) &=&t_0 -\sqrt{\rho/r_0+1}\left(r\sqrt{1+\rho/r}+(2 r_0-\rho)\arctan\sqrt{1+\rho/r}\right) \\ \nonumber
&+& r_0 \log |\frac{\sqrt{\rho+r_0}+\sqrt{\rho+r}}{\sqrt{\rho+r_0}-\sqrt{\rho+r}}|
\eea
where $\rho=r_0/((E/m)^2-1)$.  Again, we will be interested in pairs of trajectories of this form, displaced relative to each other by $\Delta t$ in the $t$ direction.  


In our analysis below, we will warm up by studying relatively boosted branes in flat spacetime before analyzing our string production problem directly in the black hole geometry.  The two analyses have some similarity, as we will see.  
Let us consider the following pair of trajectories in flat spacetime:   
\bea\label{trajectories}
Trajectory ~ 1: (v, u) &=&  (v_1,u_0) \Rightarrow x_{\pm, 1}= v_1\pm u_0, ~~~~ v_1\ge u_0 \\ \nonumber 
Trajectory ~ 2:  (v,u) &=& x_{+, 2} = e^\eta(v_1+u_0), ~~~x_{-,2}=e^{-\eta}(v_1-u_0) \\ \nonumber
\eea
Here $u_0$ is a constant, the coordinate position where the early particle (particle 1) falls into the horizon.  (For simplicity, relative to the near-horizon black hole we have made a small overall boost to take the first particle to sit at constant $u=u_0$). 
In the black hole problem, the analogue of $u_0$ is a function of $C=E/m$.
The parameter $v_1$ evolves in the solution, starting at $u_0$ at the horizon.  The boosted solution for particle $2$ enters the horizon at a 
coordinate position $x_{-,2}=0, x_{+,2}=2 e^\eta u_0$, and its trajectory is boosted by the factor $e^\eta$. 

In the black hole problem, this approximately describes the two trajectories only in the near horizon region.\footnote{One interesting subtlety we will find below is that the time variation of the velocity cannot always be ignored even in the near horizon region.}  
If we take the trajectories (\ref{trajectories}) and extrapolate them back in time, they would meet at a locus
\be\label{meeting}
x_+=\frac{2 u_0}{1+e^{-\eta}}\approx 2 u_0(1-e^{-\eta}), ~~~ x_-=-\frac{2 u_0}{1+e^\eta}\approx -2u_0e^{-\eta}
\ee  
which becomes very close to the horizon as $\eta\to\infty$.   The black hole setup does not have these entire trajectories (see figures 1 and 2), so our problem is to estimate the level of string production for the full black hole trajectories, or at least to estimate it reliably in the patch of flat spacetime which does exist in our system.

With this motivation, below in \S\ref{sec:BH}\ and \S\ref{sec:BHII}\ we will analyze string pair production for strings ending on D-particles following  relatively boosted trajectories (displaced by $\Delta t$)  in the black hole geometry.

\subsection{Organization of the rest of the paper}

The rest of this paper is organized as follows.  In \S\ref{sec:firstquantized}\ we will derive particle production from a time-dependent mass in a first quantized framework and generalize it to string theory with a time-dependent tension, recovering earlier examples in both cases.  We discuss both precise saddle point calculations and real-time production estimates.  In \S\ref{sec:boosted}\ we will use this method to calculate the production amplitude for open strings between boosted branes in flat spacetime, recovering \cite{Bachas}\cite{RelDbranes}\ appropriately.  In \S\ref{sec:BH}, we begin the analysis of open string production between D-particle trajectories with  $E<m$ in the Schwarzschild black hole.  We compute the contribution of one saddle point which
 is associated with the early brane collision that happens in the $E<m$ trajectories, and we discuss the spacetime region of non-adiabaticity as a function of $E/m$.  This effect is strongly enhanced at large boost, and provides an interesting dynamical limitation on the corresponding late-time thought experiment.  In \S\ref{sec:BHII}, we analyze trajectories for general $E/m$, focusing on a real time estimate of non-adiabaticity at horizon crossing.  First, in \S\ref{propertime}, we find that in the D-particle's infalling frame, there is strong boost-enhanced non-adiabaticity at the horizon for very small $E/m\ll \sqrt{\alpha'}/r_0$.
Additionally, working in Gullstrand - Painlev\'e coordinates which are smooth across the horizon and describe the view of auxiliary observers dropped from infinity, our real-time estimate of non-adiabaticity indicates a significant near-horizon non-adiabatic contribution for a parametrically large range of relatively boosted $E>m$ probes.  In this case the large boost between the probes does not enhance the effect, but there is a large boost between the auxiliary observers and the probe.
We also provide a Kruskal coordinate description of some aspects of the problem to facilitate comparison with the Minkowski boosted brane process.   


In \S\ref{sec:adiabatic}, we perform further checks of our method for various adiabatic cases. In \S\ref{sec:cosmo}\ we analyze the de Sitter horizon using the same method, obtaining an adiabatic answer for particles dropped in by the static patch observer.  In the final section we discuss some questions and future directions raised by this work.       

\section{First quantized calculation of particle and string production}\label{sec:firstquantized}

In this section we develop a first quantized path integral WKB method to calculate particle and string production.\footnote{These methods were developed in collaboration with J. Polchinski, and this section is based partly on his unpublished notes \cite{unpublished}.}  In the string theoretic case, this uses a saddle point approximation as in \cite{Gubser}\cite{AlbionEmil}.  It consistently incorporates effects which go beyond a na\"ive extrapolation of particle production as in the earlier examples \cite{Bachas}\cite{RelDbranes}\cite{CSGW}. As we build up this technology, we will reproduce standard earlier results on particle production, as well as a string-theoretic example which appeared earlier in \cite{CSGW}\ (where it was analyzed in a worldsheet Hamiltonian framework).  


\subsection{First-quantized particle production with a time dependent mass}\label{subsec:particle}

Let us begin by analyzing particle production in first-quantized form so as to be able to generalize to strings.
For a particle with time-dependent mass $m(t)$, the spacetime equation of motion for the corresponding field $\psi$ -- related to the Hamiltonian constraint enforcing reparameterization invariance on the particle worldline -- is
\be
(\partial_t^2 + m^2(t) +k^2)\psi(t) = 0 \,.
\ee
with $k$ the particle momentum.  In appropriate circumstances, this is amenable to a WKB solution controlled by non-adiabaticity figures of merit such as $(d\omega/dt)/\omega^2$.  

Generically, the time dependent mass leads to solutions $\psi(t)$ which mix positive and negative frequency modes as the system evolves in time, leading to particle production.  This is captured by 
\be
\frac{\langle {\rm out} | a_{\rm out}^2 | {\rm in}\rangle}{ \langle {\rm out} | {\rm in}\rangle} \,,
\ee
which we get from the positive frequency part of
\be
\frac{\langle {\rm out} | \phi(t_1) \phi(t_2) | {\rm in}\rangle}{ \langle {\rm out} | {\rm in}\rangle} \,.
\ee
The latter is equal to 
\be\label{worldlinePI}
-i \left[ \partial_t^2 + m^2(t) - i \epsilon \right]^{-1}(t_1,t_2) 
=
\int_0^\infty d\tau \int {\cal D}t \Big|_{t_1}^{t_2} \,\exp \left\{-\frac{i}{2} \int_0^\tau d\tau' \,( \dot t^2 + m^2(t)+k^2) \right\} \,.
\ee
In this section, we use $\tau$ and dot for worldline time and derivative; derivatives with respect to real time are explicitly $\partial_t$.  

The configurations $t(\tau)$ that we integrate over must satisfy the constraint
\be\label{particleconstraint}
\dot t^2 = \omega^2(t) =  m(t)^2+k^2 \,,  \label{constraint}
\ee
and the action becomes
\be\label{genwlac}
S \to \mp  \int_0^\tau d\tau' \, \omega(t) \dot  t  = \mp   \int_{t_1}^{t_2} dt \, \omega(t) \,. \label{sclass}
\ee

\begin{figure}[htbp]
\begin{center}
\includegraphics[width=10cm]{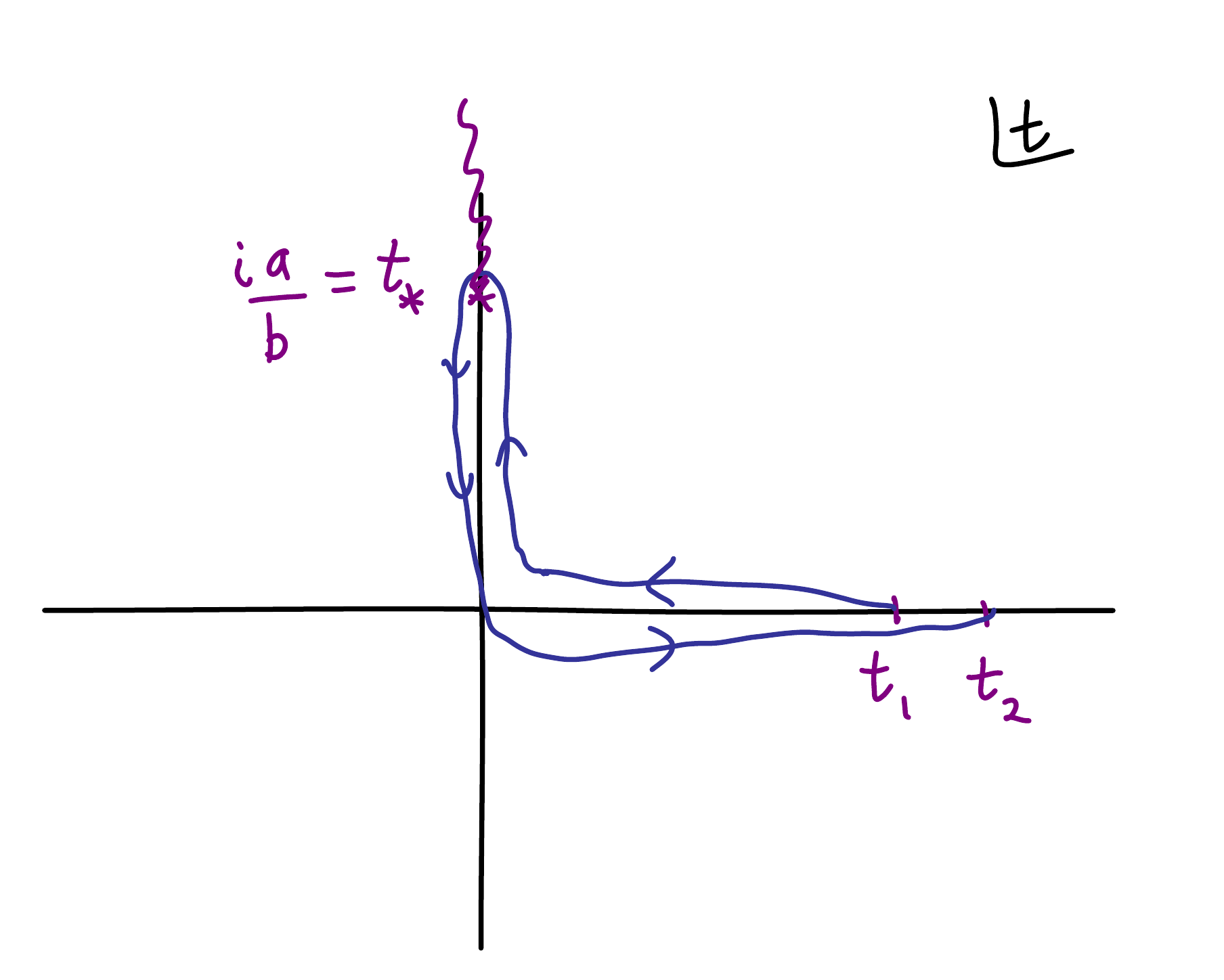}
\end{center}
\caption{The integration contour corresponding to the particle-production saddle point of the worldline path integral.  Here $t_*$ is a branch point at which the frequency vanishes, taking the value $ia/b$ in the simple example described in the text.}
\label{particlecontour}
\end{figure}

The path integral only gets contributions from paths $t(\tau)$ which satisfy the constraint (\ref{particleconstraint}). 
When both $t_1$ and $t_2$ are large and positive, there is a solution that connects them fairly directly, but this contributes to the matrix element of $a_{\rm out}^\dagger a_{\rm out}$ because  the upper and lower limits in the integral~(\ref{sclass}) contribute with opposite signs.  To obtain the $a_{out}^2$ piece, we need a different contour which approaches $t_1$ and $t_2$ in the opposite way, as in figure \ref{particlecontour}.  
The constraint (\ref{particleconstraint}) requires that a path that turns around at some value $t=t_*$ -- i.e. one that satisfies $\dot t=0$ there -- must have $\omega(t_*)=0$.  

One can think of this as a stationary phase calculation in the following sense.  Let us introduce an intermediate point $t_*$ in the contour and integrate from $t_1$ to $t_*$ and
then back up to $t_2$.  That is, take
\be\label{tstar}
S= \int_{t_1}^{t_2} dt \, \omega(t) \equiv \int_{t_1}^{t_*}dt  \, \omega(t) + \int_{t_*}^{t_2}dt  \, \omega(t)
\ee    
and consider the path integration over $dt_*$.  If we take $t_*$ such that $\omega(t_*)=0$ and integrate around the branch point at $t=t_*$ (which puts a relative sign into the second term, so the dependence on $t_*$ does not cancel), this is a stationary point of the action as a function of $t_*$ since $\frac{ dS (t_*)}{dt_*}\propto \omega(t_*)=0$.    

We can implement this in the classic example
\be\label{partab}
\omega(t)^2 = m(t)^2+k^2 = a^2 + b^2 t^2 \,.
\ee
for which the number of produced particles is given by
\be\label{mtresult}
|\beta|^2=e^{-\pi a^2/b}.
\ee
This can be calculated in a variety of ways; it reduces to a transmission/reflection problem in an analogue Schrodinger problem on an inverse Harmonic oscillator potential\footnote{This calculation is reviewed for example in  \cite{Trapping}\cite{CSGW}.}.  Our first step is to reproduce this result using our first-quantized path integral approach.  

In this theory, the relevant saddle point comes from a path that winds around the branch cut at $t_* = i a/b$.
Taking into account the extra sign from going around the branch cut, we obtain
\be\label{LLaction}
S \to -\frac{b}{2}(t_1^2 + t_2^2) - \frac{a^2}{2b}\ln\frac{4b^2 t_1 t_2}{a^2} +  \frac{i\pi a^2}{2b} \,.
\ee
Then $e^{iS}$ gives the outgoing wavefunctions times the production amplitude $e^{-\pi a^2/2b}$, reproducing the standard result (\ref{mtresult}) for the expectation value of the number of produced particles.  


Coming back to the more general case, to get the imaginary part of the action (\ref{genwlac}) we integrate vertically from the nearest point on the real axis to the branch cut, as described in e.g. \cite{Gubser}.  In general
\be
{\rm Im}\, S \approx  - 2i \int_{ t_{real}}^{t_*}  \omega(t) \,,
\ee
where the nearest singularity in $\omega(t)$ is at $t_* = t_{real} + i t_I$. 
Generically the length of the contour is of order $\omega/\frac{d\omega}{dt}$ and so the imaginary part is of order $\omega^2/\frac{d\omega}{dt}$.

\subsubsection{Real time estimates}\label{sec:realtime}

In some cases, as will occur for us specifically below in \S\ref{sec:BHII}, one would like to estimate the level of non-adiabaticity in real time, before the system has evolved through the point of minimal $\omega^2/\frac{d\omega}{dt}$.  In the problem (\ref{partab}), for example, the minimal value of $\omega^2/\frac{d\omega}{dt}$ occurs for $|t|\sim a/b$.  As we decrease $a$, the total production becomes stronger.  Suppose we are interested in production at a finite time, before the system has produced the total number of particles predicted by the above saddle point analysis.  Consider to be specific the problem  (\ref{partab}) at a time $t<-a/b$: although the minimal value of $\omega^2/\frac{d\omega}{dt}$ has not been reached, the system may already be non-adiabatic.  A plausible condition for significant non-adiabaticity is to require that $\omega^2/\frac{d\omega}{dt}$ be less than the log of the number of species of particles.  Defining the produced particle number at a finite time within a window of non-adiabaticity is not precise in general.  However, if there are many species, then for each species this estimate can still be made in a WKB regime satisfying $\frac{d\omega}{dt}\ll \omega^2$, with the sum over species contributing enough to give significant production.

For the black hole trajectories we consider, in our analysis in \S\ref{sec:BHII}, this issue will arise, and we will make an analogous estimate for real-time nonadiabaticity near the horizon.  There is a large number of species coming from the string oscillator spectrum.    In the other examples covered in this paper, we will be able to compute the production using a more precise saddle point analysis generalizing (\ref{LLaction}).  

\subsubsection{A comment on signs}

There is a small path integral subtlety when dealing with excited states.  This will not enter our most basic estimates below for stretched strings ending on D-particles, but it does arise in some more general cases involving oscillator modes and spatial momentum on higher-dimensional D-branes.  

Let us generalize to 
\be
(\partial_t^2 + m^2(t) -\partial_x^2 )\psi(t,x) = 0 \,,
\ee
adding an extra coordinate.  In a momentum eigenstate this is just replacing $m^2 \to m^2 + k^2$ as we did above.  However, the Lagrangian is now
\be
\frac{1}{2}  \,( -\dot t^2 + \dot x^2 - m^2(t)  )  \,,
\ee
and replacing $\dot x \to k$ gives the wrong sign: we have to flip it.  This can be understood as resulting from folding into the initial and final wavefunctions $e^{ikx}$.
We will also need
\be
(\partial_t^2 + m^2(t) -\partial_x^2 + x^2)\psi(t,x) = 0 \,.
\ee
Here, $-\partial_x^2 + x^2$ should be replaced by $2n+1$, where $n$ is the number of oscillator excitations (and the zero point +1 will be irrelevant for long strings).  Again to get this from the path integral we have to flip the sign of the kinetic terms.    


\subsection{First-quantized string production with a time dependent tension}\label{CircString}

As our next step, we review a case studied in \cite{CSGW}\ in which string-theoretic effects come into play, recovering the same behavior as in that work using our path integral method.  This concerns a string with time dependent tension $\mu(t)$ of the form\footnote{This arises microscopically for certain rolling axion solutions in string theory.}       
\be\label{mut}
\mu(t)^2 = a^2 + b^2 t^2.
\ee
One can consider first the simplest case of a circular string of radius $r$.  In general this changes with time.  An important string-theoretic effect arises for large enough $r$ that the variation of the string tension $\dot\mu/\mu$ is much larger than the time scale $\sim \dot r/r$ for the string to shrink.  In this regime it is not appropriate to treat string production as a sum of particle production events for the ordinary spectrum of particle states derived from oscillator levels of the string.  Instead, in this situation -- as we will see explicitly -- one obtains a frozen solution in which the time dependence $\dot r$ of the circle radius is subdominant in the dynamics.  This regime is well controlled by WKB.      

In appendix A.3 of \cite{CSGW} this was analyzed in worldsheet Hamiltonian framework, which yields a controlled WKB solution for the circular string wavefunction at large size $r$.   This leads to the following result for the expectation value of the number of pairs of strings produced at momentum $k$:\footnote{To compare to appendix A.3.1 of \cite{CSGW}, take $r_{there}=2\pi r_{here}$.}
\be\label{circbeta}
\langle N_k\rangle \sim e^{\frac{k^2}{4 b r}+\frac{\pi^2 a^2 r}{b}}.
\ee

Let us first analyze this in the worldsheet path integral framework.  The Polyakov action in conformal gauge is
\be
S = \frac12 \int d\tau\, d\sigma \, \mu(t)(-\dot t^2 + t'^2 + \dot X^2 - X'^2) \,. \label{poly}
\ee
We consider a circular string with some center of mass motion (in a direction orthogonal to the plane of the string),
\be
X = x(\tau) + r(\tau)(\hat 1 \cos \sigma + \hat 2 \sin \sigma) \,,\quad t = t(\tau)\,,
\ee
for which 
\be\label{circac}
S = \pi \int d\tau \, \mu(t) (- \dot t^2 + \dot x^2 + \dot r^2 - r^2)  \,.
\ee
The equations of motion are
\be\label{eomws}
\dot{ (\mu\dot r )} = -\mu r \,,\quad  \dot{( \mu \dot x)} = 0\,,\quad \dot t^2 = \dot x^2 + \dot r^2 + r^2 \,.
\ee
The final equation is from the constraint.
For the center of mass motion we have $2\pi \mu  \dot x = k$.

According to our prescription above, we then plug this into the action
\be\label{acI}
I =  \pi \int d\tau \, \mu (t) (- \dot t^2 - \dot x^2 - \dot r^2 - r^2) =  -2\pi \int d\tau \, \mu (t) \dot t^2 =
-2\pi \int dt \, \mu (t) \dot t  \,. \label{acti}
\ee
The exponent $B$ in the Bogoliubov coefficient $\beta\sim e^{-B}$ is twice the imaginary part of this, integrated from the real axis to the branch cut (see figure \ref{particlecontour}).

\subsubsection{Frozen approximation}

Let us first check the case $\dot r \ll r$ against the result (\ref{circbeta}) in \cite{CSGW}.  We have $\dot t^2 = r^2 + k^2/4\pi^2 \mu ^2$, and so for $\mu ^2 = a^2 + b^2 t^2$ we have
\be\label{frozenac}
B = 4\pi\, {\rm Im}  \int_0^{t_*} dt \,(r^2 a^2 + r^2 b^2 t^2 + k^2/4\pi^2)^{1/2} = \frac{\pi^2 r a^2}{b}
+ \frac{k^2}{4br} \,. \label{frozen}
\ee
Note that the position of the branch cut depends on $k$.  This agrees precisely with (\ref{circbeta}).  We will find an analogous regime in our analysis of boosted branes below; the reader eager to reach those results can skip the next two subsections.

\subsubsection{Adiabatic approximation}\label{subsubsec:adiabatic}

For completeness, let us consider now the adiabatic approximation, where the change in tension is small during one period of the string, $\dot \mu  /\mu  \ll 1$ (note that the period is $2\pi$ in the world-sheet $\tau$).  In this limit we can solve the $r$ equation via WKB,
\be
r = \rho \mu ^{-1/2} \cos \tau \,,
\ee
where $\rho$ is a constant of the motion.  In the same approximation, 
\be
\dot t^2 = \frac{\rho^2}{\mu }  +  \frac{k^2}{4\pi^2 \mu ^2}\,.
\ee
The mass of the string state is then $2\pi \rho \mu ^{1/2} = (4\pi \mu  [N+\tilde N])^{1/2}$, giving the right- plus left-moving excitation level $N + \tilde N= \pi \rho^2$.

The bounce action is now given by Eq.~(\ref{acti}).  To get an explicit expression
let $\mu ^2 = a^2 + b^2 t^2$ and set $k=0$, so
\be
B =   \frac{4\pi C \rho a^{3/2}}{b}\,,\quad C = \int_0^1 dx\,(1-x^2)^{1/4} \,.
\ee
More generally, 
\be
B = 4 \pi\rho\, {\rm Im}\int_{0}^{t_*} dt \, \mu ^{1/2}  \sim \rho \mu ^{3/2}/\partial_t \mu  \,,
\ee
 where $t_*$ is the nearest branch point to the origin.

\subsubsection{Matching}

The condition for the adiabatic approximation to be valid is
\begin{equation}
\dot \mu  /\mu  = \rho \partial_t \mu  /\mu ^{3/2}  < 1\,.  \label{adia}
\ee
 For excited strings this is stronger than both $\partial_t \mu  /\mu ^{3/2} < 1$ and $\partial_t m/m^2 \sim \partial_t \mu  / \rho \mu ^{3/2}  < 1$.  Moreover, it is exactly the same as the condition that the adiabatic-regime solution for $r$  satisfy the causality condition $\partial_t r < 1$.

For $\mu  = (a^2 + b^2t^2)^{1/2}$, the maximum of $\partial_t \mu  /\mu ^{3/2}$ is of order $ba^{-3/2}$, while for large times it is of order $b^{-1/2} t^{-3/2}$.  For given $\rho$, the adiabatic condition~(\ref{adia}) is always satisfied at large enough times.  It is satisfied at all times only if $a^{3/2}b^{-1} > \rho$, in which case $B > \rho^2$.

The adiabatic approximation always holds at large times, but for large enough $\rho$ it breaks down at small times and so we must match the two regimes: for a given adiabatic motion $r$ freezes at a value $r(\rho)$ given by
\be
r(\rho) = \rho \mu ^{-1/2} \ \ {\rm when}\ \  \rho \partial_t \mu  /\mu ^{3/2}  = 1 \,.
\ee
Generally this happens for $\mu  \approx bt$, in which case $r(\rho) = \rho^{2/3} b^{-1/3}$.  The branch cut $t_*$ lies in the frozen part of the motion, so we can insert this value for $r$ into the frozen result~(\ref{frozen}),
\be
B  = \frac{\pi^2 \rho^{2/3} a^2}{b^{4/3}}
+ \frac{ k^2}{4b^{2/3} \rho^{2/3}} \,. 
\ee
For large $\rho$ this is an enhancement over the adiabatic result, as expected.


In \cite{unpublished}\ we consider more general string configurations.  This is something worth fleshing out further, but we do not need those details for our main goal of assessing non-adiabaticity in the black hole background.  We will incorporate oscillator modes in our analysis as in \cite{RelDbranes}\ (\ref{Ntot}), where they behave similarly to the transverse impact parameter in the process.  

\section{Open String Production between Boosted Branes}\label{sec:boosted}

In this section, we reproduce the results of \cite{Bachas}\cite{RelDbranes}\ for open string production between boosted branes in flat spacetime and patches thereof, using the technique developed in the previous section.  
This is motivated both as a check of our methods and by the relation to the black hole problem. Near the horizon $r=r_0$, and working for simplicity in a locally flat region on the sphere, the black hole is approximately a Milne-like cosmology with metric
of the form
\be\label{Milnetau}
ds^2=-dT^2+T^2 dy^2+dX_\perp^2 = -dv^2+du^2+dX_\perp^2
\ee
In this section we will analyze open string production between relatively boosted branes in the flat spacetime geometry (\ref{Milnetau}), working it out in detail in both coordinate systems.  In the following sections we will analyze the black hole problem directly.


Let us compute this starting from the Nambu-Goto worldsheet action
\be\label{genwsNG}
S=-\frac{1}{\alpha'}\int d\tau d\sigma \sqrt{-{\rm det}~ G_{MN}\partial_\alpha X^M\partial_\beta X^N}
\ee
The boundary conditions at the two endpoints $\sigma=0,\sigma=\pi$ require the string to end consistently on the D0-branes which follow a generalization of the trajectories labeled 1 and 2 in the introduction (\ref{trajectories}).   For simplicity there we made an overall boost to put the first trajectory at constant $u=u_0$, but we will want to consider more general trajectories here, with
\be\label{ugeneral}
u_1 = V v_1 +u_0
\ee
The transformation of variables between $u, v$ and $Y, T$ is 
\be\label{YTuv}
u=T\sinh Y, ~~~ v= T \cosh Y.
\ee
In terms of the Milne $Y,T$ coordinates, the trajectory of the second brane is
\bea\label{YTtrajectory}
T_2 &=& \sqrt{v_1^2-u_1^2} \\ \nonumber
Y_2 &=& \eta + \log\sqrt{\frac{v_1+u_1}{v_1-u_1}}
\eea
with the trajectory parameterized by $v_1\ge u_1$.  
(In the black hole problem to be analyzed in the next section, this trajectory corresponds to that of a late-time observer
with $\eta = \Delta t/2r_0 \gg 1$ a large boost factor.) 
The trajectory of the first brane is of the same form but without the term proportional to $\eta$:  
\be\label{YTone}
T_1 =  \sqrt{v_1^2-u_1^2}, ~~~ Y_1= \log\sqrt{\frac{v_1+u_1}{v_1-u_1}}.
\ee

Let us make the ansatz 
\be\label{TYofsigmatau}
T(\sigma,\tau)=T(\tau) = \sqrt{v_1(\tau)^2-u_1(\tau)^2}, ~~~ Y(\sigma,\tau)= \eta\frac{\sigma}{\pi} + \log\sqrt{\frac{v_1(\tau)+u_1(\tau)}{v_1(\tau)-u_1(\tau)}}
\ee
where $u_1(\tau)$ is given by (\ref{ugeneral}).  
This puts the endpoints of the string on the D0-branes and stretches the string along the shift-symmetric $y$ direction.  It satisfies the correct boundary conditions at the end of the string, derived from the boundary terms in the variation of the Nambu-Goto action.  By symmetry, a saddle point within this ansatz will be a saddle point of the full worldsheet theory.  

\subsection{Analysis in Milne coordinates for three configurations: recovery of \cite{Bachas}\cite{RelDbranes}}

In this section, we will determine the level of production of pairs of open strings between the D-brane trajectories just defined.  These cut through the Rindler horizon $u=v$ in different ways, depending on the values of $u_0$ and $V$.  In the black hole problem, these are in turn determined by the time and energy to mass ratio $E/m$ of the first trajectory, with the relative boost given by $\eta=\Delta t/(2 r_0)$.  The total amount of production is the same in all cases (for a given $\eta$ and impact parameter), and we precisely reproduce the result of \cite{Bachas}\cite{RelDbranes}\ using our methods.  What depends on the parameters $u_0$ and $V$ is the overlap between the regime of non-adiabaticity -- estimated as the regime of sufficiently large $\dot\omega/\omega$ --  and the Rindler horizon $u=v$.  After calculating the total production in this subsection, we will return to this point in the following subsection.  

\subsubsection{$u_0=0=V$}

Let us first analyze the special case where $u_0=0$.  This is precisely the problem studied in \cite{Bachas}\cite{RelDbranes}.   This does not correspond to the black hole problem, for which $u_0$ is positive, growing with $R/r_0$ according to the solution  (\ref{radialfall}).  However this case will provide a simple check of our methods, enabling us to reproduce the result \cite{Bachas}\cite{RelDbranes}.                   
This corresponds to taking the first brane to lie on the trajectory $Y_1 = 0$.   
The second brane follows the trajectory $Y_2=\eta$.  We can also separate them by a distance $b_\perp$ in the $X_\perp$ directions as in \cite{Bachas}\cite{RelDbranes}.      
In that case, plugging our ansatz into the action gives
\be\label{wsacsol}
I=\int\frac{ dT}{\alpha'} \dot T=\int\frac{ dT}{\alpha'}\sqrt{T^2\eta^2 + b_\perp^2}
\ee 

Before making the contour calculation, let us estimate the degree of non-adiabaticity in our system as a function of $T$.  From (\ref{wsacsol}) we have 
$\omega(T)=\sqrt{T^2\eta^2+b_\perp^2}/\alpha'$.  The level of non-adiabaticity is roughly given by $|\beta|^2\sim e^{-\omega^2/(d\omega/dT)}$.  We have 
\be\label{nonad}
\frac{d\omega/dT}{\omega^2}=\eta \frac{(\eta T) \alpha'}{((\eta T)^2+b_\perp^2)^{3/2}}
\ee 
This is maximized at $\eta T\sim b_\perp$, leading to 
\be\label{betaest}
e^{-\omega^2/(d\omega/dT)}\sim e^{-b_\perp^2/\eta\alpha'}
\ee
up to a constant factor in the exponent not captured by this parametric estimate.  This agrees with the behavior computed in \cite{Bachas}\cite{RelDbranes}.  


Now let us calculate the production using our contour procedure, applied to the theory (\ref{wsacsol}). This has the same structure as in the above particle production case (\ref{genwlac})-(\ref{mtresult}), with the action vanishing at $T_*=ib_\perp/\eta$.  Integrating over the contour going around this branch cut and taking the imaginary part ${\rm Im} S$ of the action as above, and summing over states, we obtain an estimate for the number of strings
\be\label{Bresult}
{\rm Im} S =\frac{\pi b_\perp^2}{2\eta \alpha'}, ~~~~ \sum |\beta|^2\sim e^\eta \times e^{-2 {\rm Im} S}.
\ee
This agrees with the result of \cite{Bachas}\cite{RelDbranes}, in precisely the same regime (the factor of $e^\eta$ in the last expression coming from the density of states as explained in \cite{Bachas}\cite{RelDbranes}).\footnote{Note in our notation $b_\perp$ has dimensions of length, whereas \cite{Bachas}\cite{RelDbranes}\ define a dimensionless impact parameter by absorbing a factor of order $\sqrt{\alpha'}$.}

\subsubsection{$u_0\ne 0$, $V=0$}

Now let us analyze the trajectories with $u_0>0$.  In this case, there is not precisely a constant-$Y$ solution, as is particularly clear near the horizon $T\to 0$ where $Y\to\infty$.  

Plugging in the full ansatz (\ref{TYofsigmatau}), we obtain the action
\bea\label{YTactionone}
S &=&-\frac{\pi}{\alpha'}\int dT\sqrt{\frac{b_\perp^2}{\pi^2}\left(1-T^2\left(\frac{dY}{dT}\right)^2\right)+\left(\frac{\eta}{\pi}\right)^2T^2}\\
&=& -\frac{1}{\alpha'}\int dT T\sqrt{\frac{b_\perp^2}{T^2+u_0^2}+{\eta}^2} \\
\eea
This reduces to (\ref{wsacsol}) as it must in the case $u_0=0$



In this calculation,  we have a somewhat more involved analytic structure which we will also find in the black hole example.    There is a branch cut starting at the integrable singularity in the integrand at $T^2\equiv T_{*1}^2=-u_0^2$.   This branch cut ends at 
\be\label{TstarMilne}
T_{*2}=iu_0\sqrt{1+\frac{b_\perp^2}{u_0^2\eta^2}}
\ee
where $\omega$ vanishes.  The contour drawn in figure \ref{Tcontour}\ below (which will also arise in the direct black hole analysis) produces a nontrivial saddle point with a finite imaginary part to the action.  In this example, it agrees precisely with the previous result (\ref{Bresult}) -- the imaginary part of the action is independent of $u_0$.

The integrand also has a simple zero at $T=0$.  This naively suggests strong nonadiabaticity near $T=0$, as in the case of a particle with time dependent mass $\propto |T|$.  but the WKB approximation does not apply in this regime.  Below  we will analyze this in coordinates which smoothly cross the horizon.    

\subsubsection{The general case $V\ne 0$, $u_0\ne 0$}

Finally let us analyze the general case, in which there is an boost by velocity $V$ relative to the frame defined by the Milne patch of $Y,T$.    This is still just a pair of relatively boosted branes in Minkowski spacetime, but describes trajectories whose intersection with the $T=0$ slice of our Milne patch is similar to how the black hole trajectories cross the horizon at $r=r_0$.  (In the next section, we will analyze the black hole trajectories (\ref{radialfall}) directly.)

Analyzing this case in a similar way using the general form of the trajectory above, we obtain
\bea\label{Sboostedgen}
S &=&-\frac{\pi}{\alpha'}\int dT\sqrt{\frac{b_\perp^2}{\pi^2}\left(1-T^2\left(\frac{dY}{dT}\right)^2\right)+\left(\frac{\eta}{\pi}\right)^2T^2}\\
&=& -\frac{1}{\alpha'}\int dT T\sqrt{\frac{b_\perp^2}{T^2+u_0^2/(1-V^2)}+{\eta}^2} \\
\eea
From this we can see that this calculation reduces to the previous one, giving again
\be\label{ImSgenYT}
Im S= \frac{\pi b_\perp^2}{2\eta\alpha'}
\ee

\subsection{Regime of non-adiabaticity and near-horizon black hole probes}\label{sec:BBuv}

We have recovered the result for the total production which had been derived using the annulus diagram in \cite{Bachas}\cite{RelDbranes}.   
One question left open by this analysis so far is over what region in spacetime the non-adiabaticity is coming from, as a function of $u_0$ and $V$.  In simple particle production calculations, one can estimate this by determining the range of times over which $\dot\omega/\omega^2$ is at least of order 1 (or more precisely, its inverse needs to be at most of the order of the log of the number of species).  For example, in the single-species particle production problem $\omega^2=\mu^2 + A^2 t^2$ discussed above in \S\ref{sec:firstquantized}, we have $\dot\omega/\omega^2=A^2/(\mu^2+A^2 t^2)$, and unsuppressed production whenever this is greater than 1. 
Note that this region is not simply localized at the point $t=0$ of lightest mass; in D-brane scattering this means the non-adiabaticity is not strictly localized at the point of collision or closest approach of the branes.   

The calculation we have just made was done in Milne coordinates $Y, T$, but for nonzero $u_0$, the brane collision occurs (up to the impact parameter $b_\perp$) beyond the Milne patch, in the Rindler region outside the horizon $u=v$.     
We can write the action back in the Minkowski $u,v$ variables, having set it up as above to respect the symmetries. 
We have the first trajectory $u_1=V v_1 + u_0$, with the second trajectory shifted in the $Y$ direction as in 
(\ref{TYofsigmatau}).
This gives
\be\label{Suv}
S=-\frac{1}{\alpha'}\int dv\sqrt{b_\perp^2(1-V^2)+\eta^2[v-u V]^2}=-\frac{1}{\alpha'}\int dv\sqrt{b_\perp^2(1-V^2)+\eta^2[v(1-V^2)-u_0 V]^2}
\ee
This now takes the form of our simplest particle production example (i.e. isomorphic to the problem $\omega^2=\mu^2 + A^2 t^2$) in terms of $\tilde v=v(1-V^2)-u_0 V$.  
Despite the appearance of parameters $V, u_0$
here, the result is again precisely (\ref{Bresult}) as it should be \cite{Bachas}\cite{RelDbranes}.  However, although they drop out
of the final answer, these parameters do enter in the
regime of non-adiabaticity as a function of $v$. 

We have the non-adiabaticity estimator
\be\label{wdotMink}
\frac{\frac{d\omega}{dv}}{\omega^2}=\frac{\alpha'\eta^2(1-V^2)[v(1-V^2)-u_0V]}{((b_\perp^2+n\alpha')(1-V^2)+\eta^2[v(1-V^2)-u_0v]^2)^{3/2}} 
\ee
where we have included oscillator level $n$ as in \cite{Bachas}\cite{RelDbranes}\ (this adds to the effective transverse impact parameter $b_\perp^2$ in a simple way, something that is intuitive from the fact that oscillating strings are well approximated by transverse random walk configurations).  For large $n$, this needs to be greater than $1/\sqrt{8\pi^2 n}$ to generate large non-adiabaticity.

For our black hole problem, we will be interested in how this behaves at the Rindler horizon $u=v$, which intersects our trajectory at $u_{horizon}=u_0+Vv_{horizon}=v_{horizon} \Rightarrow v|_{horizon} = u_0/(1-V)$.  Evaluating (\ref{wdotMink}) on this locus gives
\be\label{wdotMinkhor}
\frac{\frac{d\omega}{dv}}{\omega^2}|_{horizon}=\frac{\alpha'(\eta u_0)^2\frac{(1-V^2)}{u_0}}{((b_\perp^2+n\alpha')(1-V^2)+(\eta u_0)^2)^{3/2}} 
\ee
So for example in the regime where the impact parameter and oscillators are negligible, strong non-adiabaticity at the horizon arises for 
\be\label{honadBB}
\frac{u_0^2}{\alpha'} < \frac{1-V^2}{\eta}
\ee

This result should provide a good order of magnitude estimate for the level of non-adiabaticity of general pairs of probes in the black hole problem.    However we will find in studying the Kruskal black hole that the variation of the overall velocity $V$ in the near horizon region is not always negligible, giving effects that can arise at the same order as those captured here.  In the black hole problem, the values of $u_0$ and $V$ at the horizon depend on the initial time $t$ and energy to mass ratio $E/m$ of the infalling trajectory, and the relative boost is given by $\eta=\Delta t/(2 r_0)$.  For trajectories satisfying (\ref{honadBB}), or more generally with sufficiently large (\ref{wdotMink}), the analysis here suggests significant non-adiabaticity at the horizon.  

In the next two sections we will analyze certain black hole trajectories directly, but the results of this section are a good guide for pairs of trajectories with a relative boost at the horizon, including more general black hole trajectories.

\section{Schwarzschild black hole I: full saddle point for  trajectories with $E<m$}\label{sec:BH}

With this background, let us next consider the black hole case.  To begin, we will consider radial trajectories in which the particles fall in from rest at $r=R$ at time $t=0, t=\Delta t$ respectively.  
We will calculate the production in this trajectory by evaluating a worldsheet path integral saddle point as developed in previous sections.  Below in \S\ref{propertime}\ we will make a real-time estimate of production near the horizon in an infalling proper frame.  

 We have two D0-branes that fall into a Schwarzschild black hole
\be\label{SchwarzII}
ds^2 = -(1-\frac{r_0}{r})dt^2 + \frac{dr^2}{(1-\frac{r_0}{r})}+r^2 d\Omega^2 + ds^2_\perp
\ee
at different Schwarzschild times $t$, in each case starting at some initial radius $r=R$.  
Before dropping in from $r=R$, our observer-branes each came up from the past horizon, and their trajectories meet up to a transverse impact parameter $b_\perp$ as dicussed in the introduction.  


As discussed in the introduction, the radially infalling solutions for $E<m$ take the form \cite{MTW}
\bea\label{radialfallagain}
r(\alpha) &=& \frac{R}{2}(1+\cos(\alpha))\\
t(\alpha) &=&( (\frac{R}{2}+r_0)\alpha+\frac{R}{2}\sin(\alpha))\sqrt{\frac{R}{r_0}-1}+r_0 \log\left(\left|\frac{\sqrt{\frac{R}{r_0}-1}+\tan(\alpha/2)}{\sqrt{\frac{R}{r_0}-1}-\tan(\alpha/2)}\right|\right)\\
\eea
Given the first solution, where $r(0)=R, t(0)=0$, the $t$ translation symmetry determines the second solution to simply be given by shifting $t\to t+\Delta t$.    

Let us work in Schwarzschild coordinates, with $r$ the time coordinate and $t$ a spatial coordinate inside the horizon.  
Following the procedure above, we will write down the worldsheet action for an open string stretched  between these two trajectories and see if the path integral has a saddle point with finite action contributing to the expectation value of $a_{out}^2$.  
As in the above cases, we will find that the frequency as a function of the time variable has a branch cut at a finite location on the complex plane, indicating string production.  As discussed in the introduction, the relative boost between the two trajectories as they cross the horizon (\ref{boost})(\ref{boostxpm}) implies that the string is stretched as a function of time, raising the possibility of non-adiabatic production.  However, the earlier (near-)collision of the branes contributes to the non-adiabaticity in this example, motivating a real-time estimate at horizon crossing to be performed below in \S\ref{propertime}.  

We again consider a symmetric ansatz (analagous to (\ref{TYofsigmatau}), with $\alpha=\alpha(\tau)$ and
\be\label{SchwarzWS}
t=t(\alpha(\tau)) + \Delta t \frac{\sigma}{\pi}, ~~~ r=r(\alpha(\tau)), ~~~X_\perp=b_\perp\frac{\sigma}{\pi}
\ee
This puts the endpoints of the string ($\sigma=0,\pi$) on the two D0-brane trajectories, consistently solving the boundary conditions derived from the boundary terms in the worldsheet action.  Again as in the previous example, by symmetry a saddle point in this ansatz will correspond to a saddle point of the full worldsheet path integral.   




Next, as in the above examples we would like to compute the string production amplitude by steepest descent.
As above, its magnitude is given by $Exp(-Im S)$ along an appropriate contour.  
Let us compute this starting from the Nambu-Goto worldsheet action
\be\label{genwsNG}
S=-\frac{1}{\alpha'}\int d\tau d\sigma \sqrt{-{\rm det}~ G_{MN}\partial_\alpha X^M\partial_\beta X^N}, 
\ee
which reduces to
\be\label{WSnext}
S=-\frac{1}{\alpha' \pi}\int d\tau d\sigma \sqrt{-b_\perp^2\left(-(1-r_0/r)\dot t^2+\frac{\dot r^2}{1-r_0/r}\right)+\dot r^2\Delta t^2}
\ee
where dot denotes $d/d\tau$.  

Using
\be\label{dtdr}
\frac{dt}{dr}=-\frac{\sqrt{R/r_0-1}}{(1-r_0/r)\sqrt{R/r-1}}
\ee
and integrating over $\sigma$ we can write this as
\be\label{Snice}
S=\frac{1}{\alpha'}\int dr \sqrt{\frac{b_\perp^2 (R/r_0)}{R/r-1}+\Delta t^2} \equiv \int dr \omega_r(r)
\ee
This form is very simple, and exhibits dependence on the time coordinate $r$, indicating some non-adiabaticity. 
Let us transform to interior Milne-like time coordinate $T$, defining
\be\label{Tr}
dT = -\frac{dr}{\sqrt{1-r/r_0}}, ~~~ r=r_0-\frac{T^2}{4r_0}, ~~~ T=2r_0\sqrt{1-r/r_0}
\ee
In terms of $T$ the action is 
\be\label{STSchwarz}
S = \frac{1}{\alpha'}\int \frac{dT T}{2r_0}\sqrt{\frac{b_\perp^2(R/r_0)(r_0-T^2/4r_0)}{R-r_0+T^2/4r_0}+\Delta t^2 }\sim \int dT\omega(T)
\ee

Before moving to our main calculation, let us note that if we considered the regime of parameters  and times $T$ satisfying $R-r_0\ll T^2/4r_0 \ll r_0$, this action would reduce to the simple form 
\be\label{STSchwarzsimple}
S_{simple} = \frac{1}{\alpha'}\int \frac{dT}{2}\sqrt{b_\perp^2+\frac{\Delta t^2 T^2}{r_0^2} }
\ee
which would produce the same amplitude as for relatively boosted branes in flat spacetime.  
However, we are also interested (perhaps mainly so) in the regime $R\gg r_0$.

As in \S\ref{sec:firstquantized}\ and \S\ref{sec:boosted}, we can analyze this explicitly by identifying
appropriate contours which correspond to saddle points of our worldsheet path integral.   We obtain the Bogoliubov coefficients from the exponential of the imaginary part of the worldsheet action in such a saddle point.  With our symmetric embedding of the string stretched along the $t$ direction, the relevant part of our worldsheet path integral boils down to a form analogous to  (\ref{worldlinePI}).  A contour which goes around a branch cut where $\omega=0$ solves the worldsheet constraints and is a stationary point as in the discussion of (\ref{tstar}).       
 
The frequency $\omega_r(r)$  has a branch cut between $r_{*1}=R$, with an integrable square root singularity, and the point where $\omega_r$ vanishes at 
\be\label{rstartwo}
r_{*2}=\frac{R}{1-\frac{b_\perp^2R}{\Delta t^2r_0}}.
\ee  

In the time coordinate $T$, we also have a zero of $\omega(T)$ at $T=0$ in addition to (\ref{rstartwo}).    This $T=0$ regime in Milne coordinates is not amenable to a controlled WKB description.  We will defer discussion of other contributions and the regime of non-adiabaticity to the next sections, where we will analyze
the system using proper time along the infallers, as well as performing the analysis in Gullstrand - Painlev\'e and Kruskal coordinates which are smooth across the horizon.    


In this section, we will study the nontrivial contribution from the saddle point in which the contour goes around the branch point (\ref{rstartwo}). 
The branch points $r_{*1}$ and $r_{*2}$, translated into the $T$ variable, give us a branch cut between
the following two points on the imaginary $T$ axis:
\be\label{Tbranchpoints}
T_{*1}=2ir_0\sqrt{R/r_0-1}, ~~~~T_{*2}\approx 2ir_0\sqrt{R/r_0-1}\left(1+\frac{b_\perp^2 R^2}{2\Delta t^2 r_0^2(R/r_0-1)}\right).
\ee
where in the second expression we took a large-$\Delta t$ approximation
\be\label{largeboost}
\frac{b_\perp^2 R^2}{2\Delta t^2 r_0^2(R/r_0-1)} \ll 1, 
\ee
appropriate for our interest in old black holes (fixing the other parameters).  The point $T_{*2}$ is where $\omega$ vanishes, the point around which we integrate to obtain the relevant saddle point solution.    

That is, let us apply our contour prescription of integrating up and down from the real axis at $T=0$ around the closest branch point at which $\omega=0$, the point $T_{*2}$.
The imaginary contribution to the action comes from the part of the contour going around the cut between the two points $T_{1*}$ and $T_{2*}$ (figure \ref{Tcontour}).

\begin{figure}[htbp]
\begin{center}
\includegraphics[width=8cm]{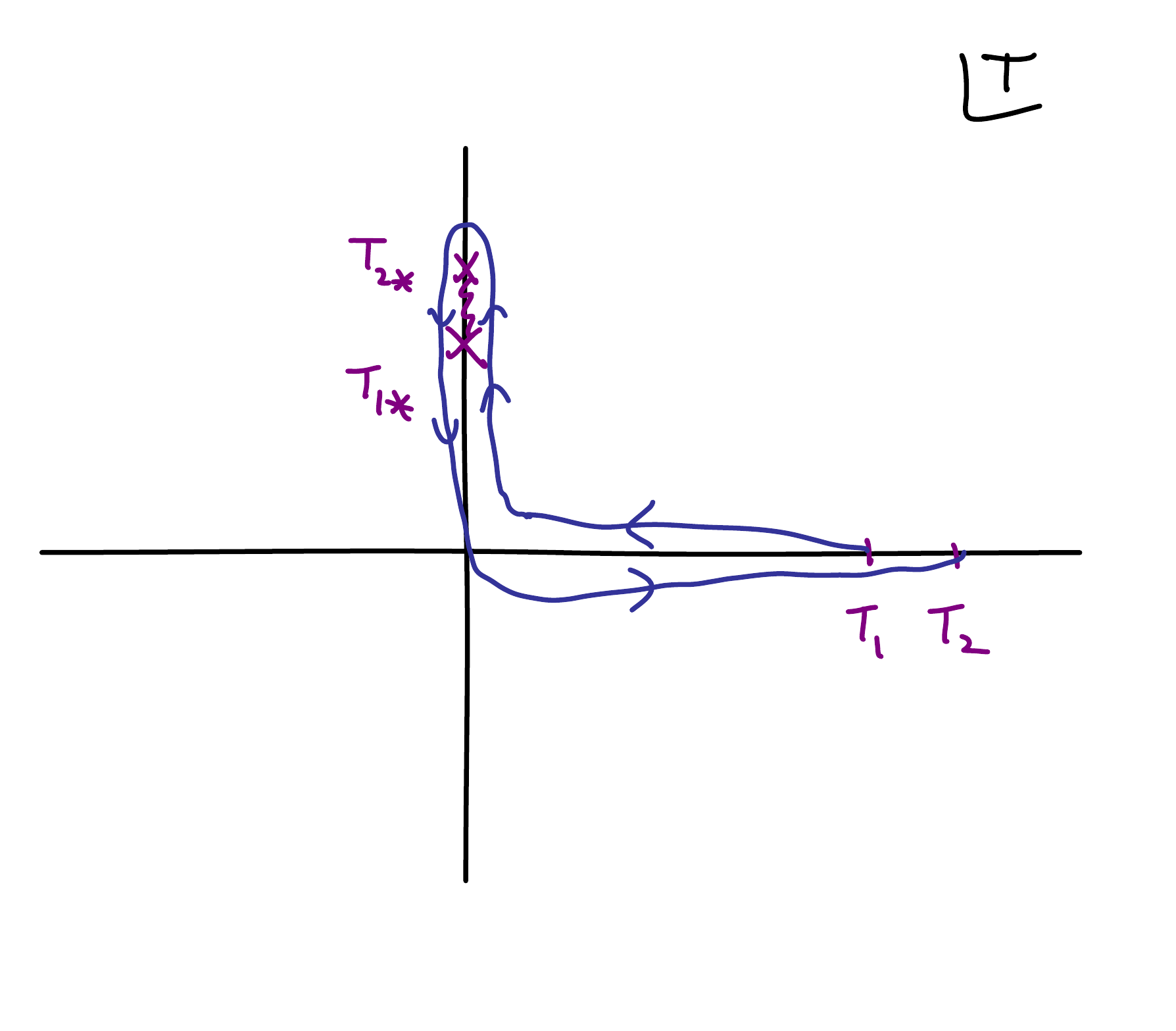}
\end{center}
\caption{The contour of integration going around the branch point where $\omega(T)=0$.  The imaginary part of the action gets a contribution from the integral around the branch cut.}
\label{Tcontour}
\end{figure}

We can calculate this explicitly in the above large-boost limit, for which $T_{*1}-T_{*1}\ll T_{*1}$.  In this regime, let us write $T=T_{*1}+\delta T=2ir_0\sqrt{R/r_0-1}+i\delta T_I$.  The integral around the cut in figure \ref{Tcontour}\ reduces to
\be\label{ImSBHone}
S_{cut} = \frac{2}{\alpha'}\int_0^{T_{*2}-T_{*1}} d\delta T_I \sqrt{R/r_0-1}\sqrt{\Delta t^2-\frac{b_\perp^2 R^2}{r_0\sqrt{R/r_0-1}\delta T_I}} 
\ee
The argument of the square root is negative in the whole range of integration, so this produces an imaginary part to the action.  After making a simple change of variables, this is given by
\be\label{ImSBHtwo}
Im S= \frac{2 b_\perp^2 R^2}{\alpha' r_0 \Delta t} \int_0^1 dw\sqrt{1/w-1}
\ee
Performing the last integral here (which is $\pi/2$), we obtain
\be\label{ImSBH}
Im S = \frac{\pi b_\perp^2 R^2}{\alpha' r_0 \Delta t} = \frac{\pi b_\perp^2 R^2}{2 r_0^2 \eta}
\ee
where in the last step we substituted $\Delta t=2 r_0\eta$.  
Again, this is valid in the regime (\ref{largeboost}); in the opposite regime with $R$ very close to $r_0$, we obtain a stronger result, similar to the boosted branes in flat spacetime from (\ref{STSchwarzsimple}).   

Note that the effect is enhanced at large $\Delta t$:  this dependence on $\eta =\Delta t/2r_0$ is as in relatively boosted branes in flat spacetime.    
At the same time, if we fixed $\eta$, the production would be suppressed at large $R/r_0$ (recall that $R$ is the radial position from which we drop in the two observers at different Schwarzschild times).  

We are ultimately interested in the production which occurs near the horizon.  
However,  the $R/r_0$ dependence in the contribution from this saddle point indicates
that the full saddle point calculation describes string production which includes the effect of the early collision between the two trajectories.
At least at large $R/r_0$ this result in fact scales correctly for string production between branes scattering in an earlier part of their trajectories (\ref{radialfall}).  They meet (up to $b_\perp$) at time $t=\Delta t/2$, after the first has fallen partly toward the black hole, and the second is still moving outward toward its turnaround point at $r=R$. 
At large $R$ for fixed $\Delta t$, their relative velocity scales like $dr/dt\sim \eta (r_0/R)^2\ll 1$.  Branes scattering at this velocity produce open strings with amplitude as in (\ref{ImSBH}).   This effect may be relevant for the setup of the thought experiment, but we are interested in extracting the part of the production which occurs near the horizon.  

We will analyze that for general $E/m$ in an infalling frame in the next section, finding a boost enhanced effect at the horizon for very small $C$.  
In any case, the non-adiabaticity contributed by this saddle point is not sharply localized at the point that the two trajectories cross; it spans a larger range in which $\omega^2/\dot\omega$ is small enough relative to the log of the density of string states.  

This calculation is at best approximate, the horizon region (and putative interior) lacking an infinite asymptotic regime in which to formulate strict S-matrix elements.  
However, our method passes a number of checks in more familiar cases, including those covered above in \S\ref{sec:firstquantized}\ and \S\ref{sec:boosted}.  
In addition, there is no significant production for a similar observer falling in early enough so as not to be strongly boosted.  Later we will show that our method reproduces the expected adiabatic behavior for particles and for strings in the absence of the relatively boosted branes, and for certain special black holes related to orbifolds of AdS.  





\subsection{Kruskal description}\label{sec:nonadiabaticity}

In this subsection, we will briefly explore the Kruskal coordinate description of our calculation.
Analogously to the boosted brane discussion above in \S\ref{sec:BBuv}, where we went from Milne to Minkowski coordinates, in our black hole analysis we can change variables from Schwarzschild to Kruskal coordinates $u,v$ (\ref{Kruskal}).  This avoids coordinate artifacts at the horizon, and facilitates comparison to the boosted brane case, with its estimate  (\ref{wdotMinkhor}) for the region of non-adiabaticity.  

For practical calculations we will find other smooth coordinates such as infalling proper time, and also Gullstrand - Painlev\'e coordinates, to be more useful, as they retain more manifest symmetry.  We will use those in the next section where we analyze general radial black hole trajectories, including those with $E>m $ which have no past collision of the D-particles.          

The worldsheet action becomes in Kruskal coordinates
\be\label{SuvBH}
S=-\frac{1}{\alpha'}\int dv (2e^{-{r/(2r_0)}})\sqrt{\frac{r_0}{r}}\sqrt{b_\perp^2(1-V^2)+\frac{\Delta t^2e^{-r/r_0}}{rr_0}[v - u V]^2}
\ee
where $V(v)=du/dv$.  Here we used the Kruskal metric (\ref{Kruskal}) and the relation
\be\label{KSdiffs}
dr=\frac{2 }{r}e^{-r/r_0}(udu-vdv)
\ee
In this action (\ref{SuvBH}) we should remember that $r=r(v)$ and $u=u(v)$ introduce more nontrivial dependence on the Kruskal time variable $v$ than was the case in the pure boosted brane example; in particular $V=du/dv$ is a nontrivial function of $v$ for the full black hole trajectory. 


One can determine the parameters at the horizon in the trajectories studied in this section, as a function of $R$.  One finds at large $R/r_0$ the behavior \cite{trajpaper}\
\be\label{largeRoverallboost}
V|_{horizon}=\frac{du}{dv}|_{horizon} = \tanh(\eta_R), ~~~ v_{horizon}=u_{horizon}\propto e^{\eta_R}~~~ \eta_R \sim \left(\frac{R}{r_0}\right)^{3/2}
\ee
where the notation $\eta_R$ refers to the overall boost of our two trajectories near the horizon as a function of $R$.  
If we send in the particles at different times (rather than dropping them at $t=0$), this effectively shifts $\eta_R$, implementing an overall boost at the horizon.   
As discussed above in \S\ref{sec:BBuv}, the resulting values of $u_0$ and $V$ determine how non-adiabatic the system is at the horizon, roughly estimated by (\ref{wdotMinkhor}).  That formula indicates regimes of significant non-adiabaticity, but strong open string pair production does not arise for every pair of trajectories.  It remains to be seen whether other effects such as discussed in \cite{RelDbranes}\cite{Veneziano}\ arise in these cases.  
   

\subsection{Dynamical limitations on thought experiments}

Before turning to other trajectories, let us make one further remark about a potential application of the trajectories studied so far with $E\le m$.  
The strong non-adiabatic behavior for a late time probe (large boost $\eta$) provides an interesting dynamical limitation on the thought experimental setup defined by these two trajectories.  If one wishes to probe the black hole early and late, using D-branes dropped in from $r=R$ as described above, this is strongly affected by open string production, which back reacts on the motion of the the probe/observers.  This, and the more dramatic Bremsstrahlung effects found in  \cite{RelDbranes}\ and closed string dynamics found in \cite{Veneziano}, may provide interesting dynamical limitations on thought-experimental setups in string theory.  We will leave this for further work, and now return to the analysis of non-adiabatic dynamics close to the horizon.     

\section{Self-Imolation in the Schwarzschild Black Hole:   general radial trajectories and non-adiabaticity at the horizon}\label{sec:BHII}

Finally, after all this warmup let us analyze radial trajectories with any value of $C\equiv E/m$.  
We will discuss both a real time estimate for $E<m$ trajectories, and also
analyze trajectories with $E>m$ which do not collide earlier in their history. 

The radially infalling solutions satisfy
\be\label{drdtEM}
\frac{dr}{dt}=-(1-\frac{r_0}{r})\sqrt{1-\frac{m^2}{E^2}(1-\frac{r_0}{r})} =-(1-\frac{r_0}{r})\sqrt{1-\frac{1}{C^2}(1-\frac{r_0}{r})} 
\ee 
Following the same steps above, separating the two trajectories in the $t$ direction and evaluating the Nambu-Goto action, this leads to a worldsheet action
\be\label{SrEm}
S=-\frac{1}{\alpha'}\int dr\sqrt{\frac{b_\perp^2}{C^2-1+r_0/r} + \Delta t^2}\equiv \int dr\omega_r(r)
\ee
For $C<1$ this reproduces the action (\ref{Snice}) of the last section, given $R=r_0/(1-C^2)$.  

\subsection{Proper time analysis and near horizon non-adiabaticity for $C\ll 1$}\label{propertime}

Let us analyze the non-adiabaticity with respect to the proper time $\hat \tau$ along the second infalling D-brane trajectory.  We find a worldsheet action 
\be\label{properws}
S=\int dr \omega_r(r) =\int d\hat\tau \frac{dr}{d\hat \tau}\omega_r(r) =\int \frac{ d\hat\tau}{\alpha'} \sqrt{b_\perp^2+\Delta t^2(C^2-1+\frac{r_0}{r})} \equiv \int d\hat \tau \hat\omega
\ee
where we used
\be\label{drdtauhat}
\frac{dr}{d\hat\tau}= -\frac{dr}{\sqrt{-(G_{tt}dt^2+G_{rr}dr^2)}}=-\sqrt{C^2-1+\frac{r_0}{r}}.
\ee
The time dependence is directly tied to the boost $\Delta t$.

In order to estimate the non-adiabaticity near the horizon, we are in need of a real-time estimate of non-adiabaticity, as in the situation described above in \S\ref{sec:realtime}.   
We will estimate the imaginary part of the action by
\be\label{Seffhat}
\frac{\hat\omega^2}{\frac{d\hat\omega}{d\hat\tau}}=\frac{\hat\omega^2}{\frac{d\hat\omega}{dr}\frac{dr}{d\hat\tau}}
\ee
Let us also now incorporate the string oscillator modes, which introduce many species of single-string states.    
As in (\ref{Ntot}) \cite{RelDbranes}, we expect the oscillator levels to contribute in a similar way as the impact parameter, captured by replacing $b_\perp^2\to b_\perp^2+n\alpha'$ for oscillator level $n$.  This is intuitive, since a massive string is generically a transversely stretched long string.
 
Implementing this,  we find
\be\label{Seffhatexplicit}
\frac{\hat\omega^2}{\frac{d\hat\omega}{d\hat\tau}}=-\frac{2 r^2 (b_\perp^2+n\alpha'+(C^2-1+(r_0/r))\Delta t^2)^{3/2}}{\alpha' r_0\Delta t^2\sqrt{C^2-1+(r_0/r)}}
\ee
At the horizon this becomes
\be\label{hSeff}
\frac{2 r_0}{\alpha'}\frac{(b_\perp^2+n\alpha'+C^2\Delta t^2)^{3/2}}{C\Delta t^2}
\ee
 That is, as above we have an estimate for the total number of string pairs produced going like
\be\label{Ntot}
N_{tot}\sim \sum_n e^{\sqrt{8\pi^2 n}-\frac{\hat\omega^2}{d\hat\omega/d\hat\tau}}\equiv \sum_n e^{K(n)}
\ee 

Let us evaluate $K$ around the peak value of $n$, and set $b_\perp=0$ for simplicity.  Specifically, we will compute it at
\be\label{nstar}
n_* = \frac{\Delta t^2\sqrt{2} \pi C}{(3 r_0\sqrt{\alpha'})}.
\ee
This gives
\bea\label{Kofn}
K(n_*) &=& \sqrt{8\pi^2 n_*}-\frac{2 r_0}{\alpha'\Delta t^2 C}(n_*\alpha'+\Delta t^2 C^2)^{3/2}\\
 &=& \frac{\Delta t}{\sqrt{r_0}{\alpha'}^{1/4}}\left(\sqrt{\frac{8\pi^3\sqrt{2}C}{3}}-\frac{2 r_0^{3/2}}{C{\alpha'}^{3/4}}(\frac{\sqrt{2}\pi C\sqrt{\alpha'}}{3 r_0}+C^2)^{3/2} \right)\\
&\approx&   \frac{\Delta t}{\sqrt{r_0}{\alpha'}^{1/4}}\left(\sqrt{\frac{8\pi^3\sqrt{2}C}{3}}-\frac{2 r_0^{3/2}}{C{\alpha'}^{3/4}}(\frac{\sqrt{2}\pi C\sqrt{\alpha'}}{3 r_0})^{3/2} \right), ~~~ C\ll\frac{\sqrt{\alpha'}}{r_0} \\
&=&   \frac{2^{11/4} \pi^{3/2}\Delta t\sqrt{C}}{3\sqrt{3}\sqrt{r_0}(\alpha')^{1/4}}, ~~~ C\ll\frac{\sqrt{\alpha'}}{r_0} \label{boostenhanced}
\eea
In the step with the $\approx$, as indicated we specialized to the case with $C\ll\sqrt{\alpha'}/r_0$.
Note that we did not need any constraint on $\Delta t$.  

The result is altogether boost enhanced, but only arises for very small $C$.  For $C<1$ the two trajectories have had an earlier collision as discussed in \S\ref{sec:BH}, but the estimate we are doing here is for production at the horizon, not at the point of the collision.    

The two members of each pair of open strings that are produced generically oscillate relative to each other, decaying into shorter closed string loops and ultimately radiation, as in \cite{CSGW}.\footnote{As in \cite{CSGW}\ we thank J. Polchinski for this observation.}  This endpoint may conform more closely to the original notion of a black hole firewall  \cite{AMPS}\ as consisting of high energy radiation, although the general arguments of \cite{AMPS}\ do not imply a specific form for the near-horizon drama they predict.  In the present example, this radiation is the decay product of an intermediate state of excited string pairs, which are in turn catalyzed by a late-time infalling observer.

\subsection{Painlev\'e time and near horizon non-adiabaticity for $C\gg 1$}

We will next analyze the horizon non-adiabaticity with respect to Gullstrand - Painlev\'e (GP) coordinates, which are smooth across the horizon and retain a manifest translation symmetry.\footnote{We thank Danjie Wenren for useful discussions of this.}  In the following subsection we will also include the Kruskal form of the action and some aspects of the trajectories, but without the manifest $t$ translation symmetry that description is less useful for practical calculations.  These coordinates correspond to the proper time not of the infalling D-branes, but of a sequence of observers dropped into the black hole from infinity.  

In GP coordinates, the metric is
\be\label{GPcoords}
ds^2_{GP}=-(1-r_0/r)dt_r^2 + 2\sqrt{\frac{r_0}{r}} dt_r dr + dr^2 + r^2 d\Omega^2
\ee
and we have
\be\label{dtrdr}
\frac{dt_r}{dr}= \frac{dt}{dr}+\frac{1}{(1-r_0/r)}\sqrt{\frac{r_0}{r}}
\ee
where $dt/dr$ is the inverse of (\ref{drdtEM}).  

In terms of the GP time coordinate $t_r$, we have action
\be\label{SGP}
S=-\frac{1}{\alpha'}\int dt_r \frac{dr}{dt_r}\sqrt{\frac{b_\perp^2}{C^2-1+r_0/r} + \Delta t^2}\equiv \int dt_r\omega_{t_r}
\ee
For the special case of $C=E/m=1$, this is straightforward to write in terms of $t_r$; we find
\be\label{SGPCone}
S|_{C=1}=-\frac{1}{\alpha'}\int dt_r\sqrt{b_\perp^2 + \Delta t^2\frac{r_0^{2/3}}{(t_{rS}-t_r)^{2/3}}}
\ee
where $t_{rS}$ is the time at which the trajectory reaches the singularity.       

In this example, we are in need of a real-time estimate of non-adiabaticity, as in the situation described above in \S\ref{sec:realtime}\ and used in the proper time analysis just above in \S\ref{propertime}.   
From our worldsheet action, the saddle point corresponding to $\omega_{t_r}(r_*)=0$ would turn around at a negative value, i.e. $r_*<0$.  This puts it out of our regime of control, but also beyond the timescale of interest for our main question about the level of non-adiabaticity near the horizon.    
Our system could have strong production near the singularity at $r=0$, but it would be difficult to control that analysis (at least aside from special cases, such as particular limits of BTZ black holes \cite{stringsing}).  
Production associated with the singularity, while interesting, is not directly relevant to the recent paradox \cite{AMPS}.

Therefore let us now simply estimate the level of non-adiabaticity at the horizon from the Painleve viewpoint, analogously to \S\ref{sec:realtime}\ and \S\ref{propertime}, by computing
\be\label{nonadGP}
\frac{\omega_{t_r}^2}{\frac{d\omega_{t_r}}{dt_r}} = \frac{\omega_{t_r}^2}{\frac{d\omega_{t_r}}{dr}\frac{dr}{dt_r}}
\ee
As we discussed above, this is a good estimate for the effective action in cases where we use the saddle point contribution to string or particle production at late times.   More generally, a real-time value of $\omega^2/\dot\omega$ smaller than the log of the number of species suggests strong non-adiabaticity, with a WKB description still possible given $\omega^2/\dot\omega \gg 1$.  
Let us again incorporate the string oscillator modes, which introduce many species of single-string states,  
by replacing $b_\perp^2\to b_\perp^2+n\alpha'$ for oscillator level $n$.  

Technically, the last form in (\ref{nonadGP}) is useful as we can directly use the expression (\ref{SGP}) for $\omega_{t_r}=\omega_r(r)dr/dt_r$ as a function of $r$.  This is straightforward to compute for all radial trajectories which depend on $C=E/m$.

Implementing this, and taking the limit $r\to r_0$ to assess the non-adiabaticity near the horizon, we find
\be\label{nonadGPresult}
\left|\frac{\omega_{t_r}^2}{\frac{d\omega_{t_r}}{dt_r}}|_{horizon}\right| = \frac{4 r_0 C^4 (1+C^2)(\frac{b_\perp^2+n\alpha'}{C^2}+\Delta t^2)^{3/2}}{\alpha'((b_\perp^2+n\alpha')(C^2-1)^2+C^2(3+C^4)\Delta t^2)}
\ee
where again $C=E/m$.  Away from the horizon, for $r>r_0$ the non-adiabaticity is weaker than (\ref{nonadGPresult}); it also grows stronger toward the singularity.\footnote{For small $C$ the non-adiabaticity is also large, something we found in the previous section.  Recall from there though that $C<1$ corresponds to pairs of trajectories which crossed (up to the impact parameter $b_\perp$) before entering the black hole.}  

At large $C\gg 1$, this boils down to
\be\label{nonadGPresultlargeC}
\left|\frac{\omega_{t_r}^2}{\frac{d\omega_{t_r}}{dt_r}}|_{horizon, C\gg1 }\right| \simeq \frac{4 r_0}{\alpha'}\sqrt{\frac{b_\perp^2 +n\alpha'}{C^2}+\Delta t^2}
\ee
This gives us a simple estimate for the imaginary part of the action for string pair production.  

Let us now put in the string theory density of states $\rho_{str}\sim Exp(\sqrt{8\pi^2 n})$, obtaining an estimate for the number of pairs of open strings produced 
\be\label{Ntotgen}
N_{tot}|_{horizon, C\gg 1}\simeq \sum_n e^{\sqrt{8\pi^2n}-\frac{4 r_0}{\alpha'}\sqrt{\frac{b_\perp^2 +n\alpha'}{C^2}+\Delta t^2}}
\ee
Thus for $C=E/m \ge {\cal O}(r_0/\sqrt{\alpha'})$ we get strong production by horizon crossing (as seen by the Painlev\'e observer).


For D-particles in string theory, we have 
\be\label{Dmass}
m=\frac{1}{g_s\sqrt{\alpha'}} = \frac{M_p}{\sqrt{\hat{Vol}}} 
\ee
where $M_p$ is the four-dimensional Planck mass and $\hat{Vol}\gg 1$ is the volume of the extra dimensions of string theory in string units.  As a result, $E/m \gg r_0/\sqrt{\alpha'}$ is consistent with $E\ll M_p$, so our infaller is a small perturbation on the background geometry.  

One notable feature of this result for the Painlev\'e non-adiabaticity estimate based on $\omega^2/\dot\omega$ is that 
evidently the production does not require a large boost between the two D-branes, which in fact suppresses the effect at large $\Delta t$ in this case.  
Since $C\gg 1$ there is, however, a large boost between the Painlev\'e observer and the infalling pair of D-branes.
Although here for simplicity in our calculation we took $X_\perp$ to be a completely transverse to the black hole geometry, a similar estimate would hold for an impact parameter along the sphere directions of the black hole geometry, which at the horizon means $b_\perp\to r_0\Delta \theta$, with $\theta$ an angular direction around the sphere.   This suggests that more generally for strings stretching along the horizon, we would obtain significant non-adiabaticity even at small $\Delta t$, for sufficiently large $C=E/m$.  The interplay among $C$, $\Delta t$, and $b_\perp$ would be interesting to understand more physically.     
Again, in the previous subsection \S\ref{propertime}\ we found boost-enhanced non-adiabaticity in the proper time of the infalling D-brane, but for small $C$ in that case.

\subsection{Kruskal description}

For some purposes, 
such as direct comparison to the Minkowski-space boosted brane problem it is interesting to change variables to Kruskal coordinates.  In this section we collect some of the relevant formulas in these variables.
This change of variables gives the worldsheet action in the form
\be\label{SruvEm}
S=-\frac{1}{\alpha'}\int dv\frac{2 }{r}e^{-r/r_0}\sqrt{(v-uV)^2 \left(\frac{b_\perp^2}{C^2-1+r_0/r}+\Delta t^2\right)}
\ee 
where $u$ is a function of $v$ along the trajectory, and $V=du/dv$.  
We can equivalently write the action in the form  (\ref{SuvBH}) discussed above,  
\be\label{SuvBHII}
S=-\frac{1}{\alpha'}\int dv (2e^{-{r/(2r_0)}})\sqrt{\frac{r_0}{r}}\sqrt{b_\perp^2(1-V^2)+\frac{\Delta t^2e^{-r/r_0}}{rr_0}[v - u V]^2},
\ee
This last form is simplest to compare physically to the boosted brane process, with action (\ref{Suv}).  
In particular, the action (\ref{SuvBHII}) reduces to (\ref{Suv}) in the near-horizon region $r-r_0\ll r_0$.  



Using the transformation between Schwarzschild and Kruskal variables, we also find the relation
\be\label{twoforms}
\frac{v-u V}{r_0\sqrt{1-V^2}}=e^{r/(2 r_0)}\sqrt{\frac{r}{r_0}}\sqrt{C^2-1+\frac{r_0}{r}} 
\ee
One subtlety in relating this Kruskal description to the boosted brane analysis is the variation of the overall velocity $V$.
To see one place where this comes in,
let us calculate the non-adiabaticity estimator
\be\label{wvnonadI}
\frac{\omega(v)^2}{\frac{d\omega}{dv}}
\ee
For simplicity, we will do so for two extreme cases:   $b_\perp=0$ and $\Delta t=0$.  
In the first case, we obtain
\be\label{wvnonadIt}
\frac{\omega(v)^2}{\frac{d\omega}{dv}} = 2 \frac{\Delta t r_0}{\alpha'}\frac{ (C^2-1+\frac{r_0}{r})}{1-\frac{u(dV/dv)}{1-V^2}+2(C^2-1+\frac{r_0}{r})}
\ee
Similarly in the second case, we obtain
\be\label{wvnonadIb}
\frac{\omega(v)^2}{\frac{d\omega}{dv}} = 2 \frac{b_\perp r_0}{\alpha'}\frac{ \sqrt{C^2-1+\frac{r_0}{r}}}{(1+\frac{r_0}{r})(2 C^2-1+\frac{r_0}{r})-\frac{u(dV/dv)}{1-V^2}}
\ee
From the denominators in (\ref{wvnonadIt}) and (\ref{wvnonadIb}) we can see that we need to understand the behavior of $dV/dv$ in order to estimate the non-adiabaticity.
After some manipulations using the formulas in the next subsection, we find
\be\label{Kfunction}
\frac{u(dV/dv)}{1-V^2} = \frac{(1-\frac{r_0^2}{r^2})u}{v\sqrt{\frac{r_0(\rho+r)}{r(\rho+r_0)}}-u}
\ee
where 
\be\label{rhofirst}
\rho = \frac{r_0}{C^2-1}
\ee
As a result, the variation of $V$ cannot always be neglected in the near horizon geometry.  

Because the proper time and Gullstrand - Painlev\'e descriptions given above are more tractable, we need not pursue this Kruskal description further here.  However, we have included it for comparison with the boosted brane analysis, a connection worth developing further taking into account the variation of $V$ as needed.

\subsection{Further details of the trajectories}

In this subsection, we collect a few details of the trajectories which are useful in some of the above analysis.
The trajectories with $E>m$ can be obtained formally by extending to $R<0$; as above we define
\be\label{rhodef}
\rho = -R = \frac{r_0}{\frac{E^2}{m^2}-1}
\ee

Integrating (\ref{drdtEM}) gives the standard solution
\bea\label{tofr}
t(r) &=& t_0 -\sqrt{\rho/r_0+1}\left(r\sqrt{1+\rho/r}+(2 r_0-\rho)\arctan\sqrt{1+\rho/r}\right) \\ \nonumber
&+& r_0 \log |\frac{\sqrt{\rho+r_0}+\sqrt{\rho+r}}{\sqrt{\rho+r_0}-\sqrt{\rho+r}}|
\eea
Changing variables to Kruskal coordinates yields the relation
\be\label{dupdum}
\frac{d(v+u)}{d(v-u)} \equiv \frac{1+V}{1-V}= e^{t/r_0}\frac{1-\sqrt{1-\frac{m^2}{E^2}(1-\frac{r_0}{r})}}{1+\sqrt{1-\frac{m^2}{E^2}(1-\frac{r_0}{r})}}
\ee
Here, as above, $V = du/dv$; this reduces in the near horizon region to the velocity of the trajectory in the frame defined by the Kruskal coordinates, the rest frame of a $u=0$ trajectory.  From this, one can solve for $V$ and obtain it as a function of $r$ in the trajectory (\ref{tofr}).  We can also determine $u$ and $v$ as a function of $r$ by plugging our trajectory  (\ref{tofr})  into the transformation (\ref{KScoords}).

\section{Adiabatic results for other cases}\label{sec:adiabatic}

In this section we verify that our methods give adiabatic results for particle production near the horizon in the Schwarzschild geometry. 
We also perform a consistency check in the case of BTZ and hyperbolic black holes, related by an orbifold to pure $AdS$ spacetime for which there is no significant production. 

As in the examples analyzed above, inside the black hole the near-horizon regime is captured by the Milne like metric (\ref{Milnetau}), and we will work with these coordinates.   
Let us consider first particle production.  Imposing the constraint and the other wordline equations of motion in the same way as in the analysis of \S\ref{sec:firstquantized}\  gives us
\be\label{particleMilne}
S=\int dT\dot T=\int dT\sqrt{k_\perp^2+m^2+\frac{k_y^2}{T^2}} = \int dT f(T)
\ee
Note that $y$ and its conjugate momentum $k_y$ are dimensionless (\ref{Milnetau}).  The only time dependence comes from the $k_y^2$ term.  This was absent for our open strings ending on D-particles localized in the $y$ direction;
in that case the leading time dependence came from the growth of the proper length of the string stretched between the boosted branes which led to the non-adiabatic behavior discussed above.   

In this example, there is a subtlety noted in e.g. \cite{Sommerfield}\cite{Milnemodes}:  the positive frequency component of the field with respect to $\alpha_T=\log(T/T_0)$ is not the same as positive frequency with respect to $T$ (or with respect to Minkowski spacetime modes).   The Minkowski vacuum is formally an excited state in the Fock space constructed with respect to $\alpha_T$.  In (\ref{particleMilne}) there is a pole in $f(T)$,  which dominates the behavior near $T=0$, giving $S\sim \int d\alpha_T k_y$, so it seems that our calculation is picking out positive frequency modes with respect to $\alpha_T$.     

In the presence of transverse momentum $k_\perp$ we have the full worldline action
\be\label{Salpha}
S=\int d\alpha_T\sqrt{k_y^2+e^{2\alpha_T} k_\perp^2}=\int d\alpha_T\omega(\alpha_T)
\ee
There is a zero of $\omega(\alpha_T)$ in the upper half plane at $\alpha_*= i\pi/2 + \log(k_y/k_\perp)$. Integrating around this gives an action with imaginary part $\pi k_y$ (one gets the same result using the $T$ variable taking into account the integral partway around the pole at $T=0$).  This agrees with the Bogoliubov coefficient that captures the change of basis between the two sets of modes in equation  (2.28) of \cite{Sommerfield}.        
Therefore our calculation appears to be consistent with the Minkowski vacuum (which does not behave like the vacuum in terms of the modes defined with respect to $\alpha_T$).    


This analysis was for particles in Milne spacetime, which is simply Minkowski spacetime.  We have recovered the fact that there is no particle production in that situation, having identified the path integral calculation as picking out positive frequency modes with respect to $\alpha_T$.  In the full black hole geometry, there are deviations from the Milne-like metric which shift the branch cut away from the imaginary axis.  This residual time dependence leads to very small non-adiabaticity.


\subsection{BTZ and hyperbolic black holes on the Coulomb branch}

In this section, we analyze yet another test case for our analysis, that of black holes with D-branes which are orbifold projections of $AdS/CFT$ on its moduli space \cite{BTZ}\cite{emparanhyp}\cite{InsightfulD}.  

Before including the D-branes, these black holes\footnote{for the particular energy level $\mu=0$ in the notation \cite{emparanhyp}}\ are projections of a region of the Poincare patch.  The metric is
\bea\label{hypmett}
\frac{ds^2}{\ell^2} &=& \frac{-dt_p^2+t_p^2 ds^2_{H_n}+dz_p^2}{z_p^2}+ds^2_{\perp}\\
&=&  \frac{-d\tilde t_p^2+ d\vec x^2+dz_p^2}{z_p^2}+ds^2_{\perp}\\
&=& -(\frac{r^2}{\ell^2}-1)dt^2 + \frac{dr^2}{\frac{r^2}{\ell^2}-1}+ r^2 ds_{H_n}^2
\eea 
where $ds^2_{H_n}$ is the metric along $n$-dimensional hyperbolic spatial slices (for BTZ \cite{BTZ}, $n=1$ and this is just the real line).
This theory has a time translation symmetry $\tilde t_p\to \tilde t_p+constant$ and should generate no production for strings stretched between the branes. 

Using the change of variables (inside the horizon) 
\be\label{ps}
t_p=-\frac{r\ell}{\sqrt{\ell^2-r^2}}e^{-t/\ell}, ~~~ z_p=\frac{\ell^2}{\sqrt{\ell^2-r^2}}e^{-t/\ell},
\ee
we see that shifting $t$ to $t+\Delta t$ rescales $z_p$, mapping a trajectory at constant $z_p=z_{p1}$ to a trajectory at constant $z_p=z_{p2}=z_{p1}e^{-\Delta t/\ell}$.  Let us denote $\eta=\Delta t/\ell$.   On the Coulomb branch,  we have one D-brane at constant $z_p=z_{p1}$ and the other at $z_p=z_{p2}=e^{-\eta}z_{p1}$.    A string worldsheet stretched between the two trajectories is described by an embedding as in the Schwarzschild black hole case analyzed above: 
\be\label{BTZansatz}
t=t(\alpha(\tau)) + \Delta t \frac{\sigma}{\pi}, ~~~ r=r(\alpha(\tau)), ~~~X_\perp=b_\perp\frac{\sigma}{\pi}
\ee
Plugging this into the Nambu-Goto action, we obtain
\be\label{Swszu}
S =\frac{1}{\alpha'} \int dr \sqrt{\ell^2\eta^2+b_\perp^2} 
\ee
where we used $(dt/dr)^2=(r^2/\ell^2)/(r^2/\ell^2-1)^2$ on the trajectory of constant $z_p$, obtained
using the transformation (\ref{ps}) and  $dz_p=0$.  This constant frequency leads to adiabatic behavior.  

With this form of the action, we can also change variables to coordinates which extend across the horizon, such as $t_p,z_p$.  In our simple trajectory with $dz_p=0$, $dr=dt_p \frac{\ell}{z_p}$ with $z_p$ constant.  So again in the $t_p$ coordinates, there is no time dependence and the system is adiabatic.  

This is very different from the Schwarzschild case analyzed above, even though both cases involve a boost transformation.   In particular, we get no imaginary part from a  branch cut  like we do there.    

This is a useful check of our methods, since the physics is manifestly adiabatic in the covering space of these BTZ and hyperbolic black holes.  Now if we make the orbifold projection $H_n\to H_n/\Gamma$ to produce the black holes constructed in \cite{BTZ}\cite{emparanhyp}, we will obtain residual non-adiabatic effects for particles and strings which sense the contraction of the compact space $H_n/\Gamma$.  This includes momentum modes and wound strings.  


These black holes are special theoretical laboratories for the information problem, with much longer time-scales for evaporation and a connection to gauge theory moduli space dynamics \cite{InsightfulD}.  It would be interesting to check if the D-branes by which they decay or their antibrane partners provide relatively boosted endpoints for strings which could be produced significantly.  The analysis of various aspects of these cases remains in progress \cite{InsightfulD} \cite{RamanBTZ}.

\section{The de Sitter horizon}\label{sec:cosmo}

In cosmology, there is strong evidence for the re-entry of super-horizon perturbations (as generated in inflation)  providing the seeds for the observed structure in the universe.  There is no immediate AMPS paradox for cosmology, and no direct analogue of the black-hole formation and evaporation process.   But it is of interest to understand if the firewall dynamics we derived above applies also in this case and if so what constraints it places on our theory or its applicability in the observed universe.    

In this section, we explore the analogous calculation for cosmological horizons.  Specifically, consider an observer patch of de Sitter spacetime
\be\label{staticpatch}
ds^2=-(1-\frac{r^2}{L^2})dt^2+\frac{dr^2}{1-\frac{r^2}{L^2}}+r^2 d\Omega^2
\ee
The observer sits at $r=0$ and the horizon is at $r=L$.  

As in the black hole analysis above, let us send in two D0-brane observers separated by a static patch time translation $t_2=t_1+\Delta t$.  First let us derive their trajectories.  A particle of mass $m$ (which is $\sim 1/g_s\sqrt{\alpha'}$ for the D0-brane) has action  
\be\label{SpartdS}
S_{part}=-m\int d\tau\sqrt{(1-r^2/L^2)\dot t^2+\dot r^2/(1-r^2/L^2)}
\ee
from which one can derive the conjugate momentum and Hamiltonian, leading to conserved energy
\be\label{EdS}
E=\frac{m\sqrt{1-r^2/L^2}}{\sqrt{1-\frac{(dr/dt)^2}{(1-r^2/L^2)^2}}}
\ee
and hence
\be\label{drdtdS}
\left(\frac{dr}{dt}\right)^2=(1-\frac{r^2}{L^2})^2(1-\frac{m^2}{E^2}(1-\frac{r^2}{L^2}))
\ee
within the static patch.  This corresponds to a conserved momentum on the other side of the horizon, where $t$ becomes a spatial variable.  

Let us analyze this for the case that the particle starts at rest at $r=0$, the position of the observer
defining our static patch.   This means $E=m$.
Then integrating (\ref{drdtdS}), we find the open string configuration analogous to (\ref{SchwarzWS}) to be
\be\label{dSWS}
t=\Delta t\frac{\sigma}{\pi}+L \log \left|\frac{r(\tau)}{\sqrt{r(\tau)^2-L^2}}\right|, ~~~ r=r(\tau)
\ee
whose endpoints at $\sigma=0,\pi$ lie on the two D0-brane trajectories.

Working out the worldsheet Nambu-Goto action, we find
\be\label{dSNG}
S=-\frac{1}{\alpha'\pi}\int dr\sqrt{\frac{b_\perp^2 L^2}{r^2}+\Delta t^2}
\ee
This is isomorphic to the form we found above in (\ref{particleMilne}) for a particle in the future Milne region of pure Minkowski space, which is adiabatic.  It would be interesting to study this further, including other trajectories in the de Sitter or more general FRW observer patch.  


   
\section{Concluding comments and questions}

In this work we investigated string creation effects, raising the possibility that generalizations of \cite{Bachas}\cite{RelDbranes}\cite{CSGW}\cite{Veneziano}, some of which arise in the presence of large boosts, may play a role in black hole thought experiments.  In the black hole context, late-time observers develop strong boosts as they cross the horizon, relative to early observers and formation matter.  
Because string-theoretic non-adiabatic effects can in some circumstances be strongly enhanced relative to naive extrapolations of effective field theory \cite{Bachas}\cite{RelDbranes}\cite{CSGW}\cite{Veneziano}, and the AMPS paradox suggests that the effective field theory vacuum does not survive, it is important to assess the level of non-adiabaticity near the black hole horizon in string theory. 

This non-adiabaticity may be catalyzed by the late-time observer itself.  Focusing on open string production between D-particles, we found significant non-adiabaticity for two families of relatively boosted probes of the Schwarzschild black hole.  For $E<m$, the open string production is strongly boost enhanced but partly associated with a prior interaction;
this latter effect may provide an interesting dynamical limitation on thought experiments involving these observer trajectories.  Moreover, for $E\ll m\sqrt{\alpha'}/r_0$, we find strong boost-enhanced non-adiabaticity at the horizon, in the proper infalling frame of the D-particle (\ref{boostenhanced}).  
For $E>m$ there is no early interaction.  There we find a different regime of strong non-adiabaticity near the horizon from the point of view of auxiliary observers whose proper time is the Painlev\'e time.  This has a more complicated dependence on $\eta=\Delta t/2r_0$.   According to our estimate (\ref{nonadGPresult}) and (for $E\gg m$ (\ref{Ntotgen})), this does not require a large boost between the branes, although there is a large boost between the infallers and the auxiliary observers in this case.  For these estimates of horizon non-adiabaticity in the black hole trajectories, we used the real-time behavior of the non-adiabaticity estimator $\omega^2/\dot\omega$ near the horizon rather than a complete saddle point calculation.  This was because the only precise saddle point appears to capture effects including either a prior collision of trajectories, or the regime of the singularity rather than just the near-horizon region. 
 
The first quantized path integral method we developed passes a number of precise checks in known cases including intrinsically string-theoretic systems such as \cite{Bachas}\cite{RelDbranes}\cite{CSGW}, where non-adiabaticity
is stronger than in naive extrapolations of particle production rates in effective field theory.  Although we were motivated in part by black hole physics, this method may have wider applicability in other contexts where one requires a reliable estimate for string-theoretic nonadiabatic effects.  Although it is not arbitrarily precise within the finite timescales available in the black hole problem, we expect that our method provides reasonable estimates.        





As we said above in \S\ref{sec:BHII}, this excitation above the vacuum state ultimately leads to radiation of closed strings and gravitons. 
The two members of each pair of open strings that are produced generically oscillate relative to each other, decaying into shorter closed string loops and ultimately radiation, as in \cite{CSGW}. This endpoint may conform more closely to the original notion of a black hole firewall  \cite{AMPS}\ as consisting of high energy radiation, although the general arguments of \cite{AMPS}\ do not imply a specific form for the near-horizon drama they predict.  In the present example, this radiation is the decay product of an intermediate state of excited open string pairs, which are in turn catalyzed by an infalling observer.     

This dynamical thought experiment can be formulated more generally as an approach to the AMPS paradox \cite{AMPS},
with other probes and effects yet to be analyzed.
Here we analyzed arguably the simplest kind of probe in string theory, namely D-branes.\footnote{Another recent work \cite{berenstein}\ made a different proposal for a firewall involving the dynamics of a D-brane probe in the D0-brane black holes studied numerically there, while e.g. \cite{InsightfulD}\ and \cite{wiseman}\ come to a somewhat different conclusion about the role of the off-diagonal modes in the matrix theory of D-brane black holes; this remains an interesting direction.} 
It would be very interesting -- and important for ultimately determining the leading non-adiabatic effect for general probes of black holes in string theory -- to relax the assumption that observers are made of non-recoiling D-branes, and analyze additional D-brane effects as well as other types of probes such as massive closed strings, or even wave packets of massless fields.  These cases may require an analysis that is higher-order in the string coupling in order to carefully capture the effect of the relative boost.  Even in the D-brane case, there could be new effects at higher orders which depend differently on the boost parameter $\eta$.  Effects such as those in \cite{Veneziano}\ and the second reference in \cite{RelDbranes}\ which arise at large boost could play an important role.\footnote{It is a pleasure to thank
Thomas Bachlechner, Liam McAllister, Gabriele Veneziano, Danjie Wenren, and Matthew Dodelson for preliminary discussions along these lines.}              

The motivating idea in this work was that the large relative boost that arises for a late-time probe of the black hole may ultimately be responsible for the breakdown of effective field theory -- suggesting a `trans-Planckian solution'.  
The annulus diagram connecting our two D-branes contains closed string processes which are subject to large blueshifting as modes propagate between the relatively boosted branes.  Our results display a nontrivial boost dependence, in some cases indeed enhancing production (\ref{boostenhanced}), but not universally; our results depend nontrivially on $E/m$ and on the observer frame as well.  It would be interesting to investigate further the physical interpretation of the results.\footnote{We thank D. Marolf for discussions.}
  

For these string-theoretic processes, although effective field theory requires completion and can underestimates the nonadiabaticity, it is replaced by perturbative string theory in a relatively standard way.\footnote{at least standard relative to modifications of quantum mechanics, or ad hoc deviations from effective field theory, although these ideas have also been fruitful to explore.  Previous recent approaches to black hole problems include instructive toy models for unitarity \cite{mathurtheorem}\cite{giddings}, AdS/CFT \cite{InsightfulD}\cite{PRetc}\cite{MP}\ and other approaches to emergent spacetime \cite{banks}, quantum information constraints \cite{Qinfo}\ and the relation between entanglement and geometry \cite{EPR}.  Additional general arguments in favor of firewalls appeared in e.g. \cite{BoussoFW}. }    
Given the non-adiabatic effect described here and appropriate generalizations, it will be interesting to understand if black hole timescales such as \cite{Pagetime}\cite{scrambling}\ enter into the dynamics. 





\section*{Acknowledgements}

I am particularly grateful to J. Polchinski for collaboration on many aspects of this, starting with \S\ref{sec:firstquantized}\ and \cite{CSGW}, as well as detailed discussion of this application of our methods to black holes.  This work is dedicated to him on the occasion of his 60th birthday celebration.  This work also benefited greatly from extensive collaboration and discussion of related topics with A. Lawrence, G. Horowitz, S. Shenker, and D. Wenren.  It is also a pleasure to thank T. Bachlechner and L. McAllister for discussions of string production and Bremsstrahlung related to \cite{RelDbranes}\cite{CSGW}, and I am grateful to M. Dodelson and G. Veneziano for preliminary discussions of string probes.  I thank D. Stanford and L. Susskind for helpful explanations of the `trans-Planckian problem' which helped stimulate the final phase of this work.  I am grateful to A. Almheiri, R. Bousso, S. Gubser, L. McAllister, J. Polchinski, J. Sully, and D. Wenren for useful comments on a draft.  
It is a pleasure to thank X. Dong, S. Giddings,  D. Harlow, S. Kachru,  J. Maldacena, D. Marolf, D. Page, H. Verlinde, and many other participants, and organizers, of workshops on this subject including the 2012 KITP program ``Bits, Branes, and Black Holes", the fall 2012 Stanford firewall meeting, the CERN theory institute on black hole physics,  the KITP program ``Black Holes: Complementarity, Fuzz, or Fire", and the fall 2013 PCTS workshop  ``Bulk Microscopy from Holography and Quantum Information".  This research was supported in part by the National Science Foundation under grant PHY-0756174 and NSF PHY11-25915 and by the Department of Energy under contract DE-AC03-76SF00515.


\end{document}